\documentclass[a4paper,11pt]{article}
\usepackage{jheppub} % for details on the use of the package, please see the JINST-author-manual
\usepackage{lineno}
\usepackage[caption=false]{subfig}

\DeclareMathOperator{\csch}{csch}

\def\red{\color{red}}

\def\rar{\rightarrow}

\def\ra{\rangle}

\def\no{\nonumber}
\def\bea{\begin{eqnarray}}
\def\eea{\end{eqnarray}}
\def\be{\begin{equation}}
\def\ee{\end{equation}}

\arxivnumber{1234.56789} % if you have one

\title{\boldmath Quantum Fisher information of a cosmic qubit undergoing non-Markovian de Sitter evolution}

\author[a]{Langxuan Chen}
\author[a,b]{Jun Feng\footnote{Corresponding author}}
\affiliation[a]{School of Physics, Xian Jiaotong University, Xi'an 710049, Shaanxi, P.R. China}
\affiliation[b]{Institute of Theoretical Physics, Xian Jiaotong University, Xi'an 710049, Shaanxi, P.R. China}

\emailAdd{j.feng@xjtu.edu.cn}

\abstract{We revisit the problem of the thermalization process for an Unruh-DeWitt (UDW) detector in de Sitter space. We derive the complete dynamics of the detector in the context of an open quantum system, without utilizing Markovian or RWA approximations. We employ quantum Fisher information (QFI) to estimate the Hubble parameter, which serves as a process function to distinguish the thermalization paths in the detector's Hilbert space, determined by its local properties, including the detector's energy gap and its initial state preparation, or by the global spacetime geometry. We find that the non-Markovian contribution generally reduces the QFI compared to the Markovian approximated solution. Regarding arbitrary initial states, the late-time QFI converges to an asymptotic value. In particular, we are interested in the background field within the one-parameter family of $\alpha$-vacua in de Sitter space. We show that for general choices of $\alpha$-vacuum, the asymptotic values of the converged QFI are significantly suppressed compared to previously known results for the Bunch-Davies vacuum. }

\begin{document}
\maketitle
\flushbottom
\section{Introduction}
\label{sec:intro}

The quantum field theory in de Sitter space has been a recurrent concern for several decades. This is partly because de Sitter space is an analytically tractable curved solution to Einstein's equation with maximal symmetry. Additionally, it plays a key role in modern cosmology, as the exponentially expanding de Sitter patch predicts the evolution of the universe during cosmic inflation and in the distant future when dark energy dominates. However, the exact behavior of quantum fields in de Sitter space remains very elusive \cite{DS1}. For example, even neglecting issues of back-reaction, the QFT in de Sitter becomes catastrophic from the view of local observers, e.g., the usual S-matrix intuition of local quantum field theory seems to fail \cite{dS1}. 

An important tool for unveiling the properties of quantum fields within specific spacetime geometries is the Unruh-DeWitt (UDW) detector \cite{UDW1,UDW2}, which is a microscopic two-level quantum system (qubit) coupled locally to fluctuating backgrounds. From the perspective of detector-field interaction, the response function for a UDW detector represents the rate of quantum transitions occurring per unit proper time, determined by the quantum field propagators with respect to the interested spacetime geometry \cite{UDW3}. For example, for an accelerated UDW detector in flat spacetime, the periodic correlation functions in Rindler geometry lead to the celebrated Unruh radiation locally perceived by the detector \cite{UDW4}. Outside a static black hole, relative to a Hartle-Hawking vacuum \cite{UDW5}, the detector would be excited to equilibrium with a Planck-distributed spectrum, referred to as the Hawking effect \cite{UDW6}. Returning to the de Sitter context, it has been shown \cite{dS12-1} that a comoving UDW detector will perceive radiation with a thermal spectrum of temperature $T_H=H/2\pi$, where $H$ denotes the Hubble parameter. This so-called Gibbons-Hawking effect is quite universal for various {couplings to curvature \cite{dS12-2} and different types of quantum fields}.

Essentially, what the detector-field method provides is a local manifestation of the so-called thermalization theorem in curved spacetime \cite{UDW7}. This method determines the detector's equilibrium with background fields, representing a unique thermalization \emph{end}, whose thermal nature can be justified by the detailed-balance condition or the Kubo-Martin-Schwinger (KMS) condition of the response function \cite{UDW8}. However, this approach cannot capture the full dynamics of the detector; that is, we cannot know about how the off-diagonal terms (coherence) of the detector's density matrix evolve during the thermalization \emph{process}. Indeed, achieving the same equilibrium, characterized by a specific temperature, does not mean \cite{UDW9} that the detector must evolve along the same thermalization path in Hilbert space. 

To fully explore the dynamic nature of quantum fields in curved spacetimes, an alternative perspective of open quantum systems has been extensively utilized in recent years, particularly for determining the complete dynamics of UDW detectors in various backgrounds \cite{Open1-1,Open1-2,Open1-3,Open1-4,Open1-5}. In this context, the UDW detector is regarded as a local open system, while background quantum fluctuations serve as the environment that induces dissipation and decoherence. This method is particularly effective for tracking the long-time processes of UDW detectors, such as thermalization undergone Hawking-Unruh or Gibbons-Hawking effects. 

Once knowing the dynamics of a detector, the pressing issue is naturally to identify certain \emph{process functions} that can characterize the time-evolution details of open processes. Beyond response functions, many significant feature functions have been proposed, such as the time-dependent entropic uncertainty bound \cite{Open2,Open3}, the geometric phase of the detector \cite{Open4,Open5}, as well as the quantum coherence, which is a critical resource for the non-equilibrium entropy production of the UDW detector during thermalization \cite{Open6}.

In this paper, we revisit the dynamics of a UDW detector in de Sitter space within the context of an open quantum system. Accompanying many previous studies \cite{Open1-3,dS12,dS13,dS14}, we take further steps based on the following considerations:

Firstly, to obtain a resolvable semigroup master equation for the UDW detector, several assumptions are typically made \cite{Open1}, namely, (1) the {Born approximation}, which assumes a weak coupling between the system and environment, (2) the {Markov approximation}, which leads to memoryless evolution of the system resulting in a time-local master equation, and (3) the {rotating wave approximation} (RWA), which neglects all rapidly oscillating terms near system resonance. However, it has been argued in various ways that these assumptions may only be valid within a rather restricted parameter space \cite{dS15-1}, and more critically, the key Markov approximation may turn out to be overly drastic for the early-time dynamics of the UDW detector \cite{dS15-2,dS15-3}. Thus, to fully explore the dynamics of the UDW detector at both early and late times, the non-Markovian contributions of its open process must be taken into account. 

Secondly, the QFT in de Sitter is complicated by the existence of infinite vacua consistent with CPT invariance, labeled by a superselection parameter $\alpha$ \cite{dS2,dS3}. Among this family of vacua, the well-known Bunch-Davies vacuum (with $\alpha\rar-\infty$) uniquely extrapolates to the Minkowski vacuum in the limit of a vanishing cosmological constant, while other vacua can be formally realized as squeezed states over the Bunch-Davies vacuum. Understanding whether interacting field theory in any $\alpha$-vacua is a consistent theory \cite{dS4,dS5,dS6,dS7} and how this can provide a better understanding of new physics at the Planck scale of the early universe \cite{dS8,dS9,dS8+} or even in holographic scenarios \cite{dS10,dS11} remains a matter of debate. Therefore, in line with previous studies of the UDW detector model in the Bunch-Davies vacuum, we aim to extend the analysis to a model within the general $\alpha$-vacua.

In this study, we investigate a comoving UDW detector in the $(3+1)$-dimensional de Sitter spacetime, modeled by a two-level system that interacts weakly with massless scalar fields. The structure of the paper is as follows:

In Section \ref{sec:Master Equation}, we derive the master equation of the detector. While most previous analyses rely on the Markov approximation to simplify the calculation, we do not appeal to either Markov or RWA approximations. The solution of the equation and its dynamical features for de Sitter backgrounds are analyzed in Section \ref{Solving the Dynamics of the Qubit}, taking into account the delicate choice of $\alpha-$vacua for the background correlation functions given in Section \ref{sec:corrlation}. Our work extends earlier studies \cite{dS15-2,dS15-3} examining the non-Markovian evolution of a uniformly accelerating qubit in flat spacetime, but differs in that it includes the detailed calculations of the residues of the inverse Laplace-transformed density matrix elements in de Sitter space. This improvement is necessary to prevent the solutions from diverging logarithmically and to ensure they match the initial conditions as $\tau\rightarrow 0^{+}$. 

After obtaining the non-Markovian dynamics of the detector with respect to de Sitter $\alpha$-vacua in Section \ref{sec:QFI}, we utilize quantum Fisher information (QFI) as a process function to distinguish thermalization paths of the comoving UDW detector, which undergoes Gibbons-Hawking radiation from the cosmological horizon. This is because QFI, as an operational measure of the distinguishability of quantum states, can discriminate parameterized states $\rho(X)$ and $\rho(X+\delta X)$ with an infinitesimal change in the parameter $X$ (as summarized in Section \ref{sec:Def_QFI}). Extensive work has been done utilizing QFI to probe extremely sensitive quantum gravity effects \cite{Open8-1,Open8-2,Open8-3} or spacetime structure \cite{Open8-4,Open8-6,Open8-7}. 

In Section \ref{QFIindS}, we argue that with $H$ as a parameter chosen to be metrologically estimated, given the explicit dynamics of a comoving UDW detector, the time-evolving QFI $\mathcal{F}_Q(H)$ can distinguish the non-Markovian dynamics from those inherited from the Markov approximation. We find that after a sufficiently long time, the $\mathcal{F}_Q(H)$ approaches an asymptotic value \cite{Open8-5}, which is irrelevant to the detector's initial state preparation. However, with varying Hubble parameters or choices of de Sitter vacuum, the asymptotic value of the QFI and the ways the detector approaches it will be modified.

Throughout the paper, we use units with $G=c=\hbar=k_B=1$.

\section{The non-Markovian evolution in de Sitter space}

\subsection{Model setting}
\label{sec:Master Equation}
The UDW detector is modeled by an idealized point particle with two internal energy levels, described by a $2\times 2$ Hamiltonian 
\begin{equation}
\label{d-Hamiltonian}
    H_{S}=\frac{\omega}{2}(\vec{n}\cdot \vec{\sigma}),
\end{equation}
where $\sigma_{i}$ are the Pauli matrices, $\vec{n}$ is a unit vector, and $\omega$ is the energy gap of the two level system. %Without loss of generality, one may take $\vec{n}=\hat{z}$ when it is convenient. 

For the detector moving along a particular trajectory $x(\tau)$ in the given spacetime $\mathcal{M}$, we consider it weakly interacts with some massless scalar fields $\phi_{\alpha}(x)$ through the Hamiltonian
\begin{equation}
    H_{I}(\tau)=g\sum_{i=1}^{3}\sigma^{i}(\tau)\otimes \Phi_{i}(\tau),
    \label{HI}
\end{equation}
where $\Phi_{i}(\tau)= \sum_\alpha \chi_{i}^{\alpha}\phi_{\alpha}(x(\tau))$ are linear combinations of field operators, $g$ is a small coupling and $\sigma^{i}(\tau)$ is generated by a time translations $\mathcal{U}_{\tau}$ as $\sigma^{i}(\tau)\equiv \mathcal{U}_{\tau}[\sigma^{i}]=e^{iH_S \tau}\sigma^{i}e^{-iH_S \tau}$. 

In the interaction picture, the evolution of the density matrix $\rho_{\text{tot}}$ of the combined system is governed by 
\begin{equation}
    \dot{\rho}_{\text{tot}}(\tau)=-i[H_I(\tau),\rho_{\text{tot}}(\tau)].
\end{equation}
In the weak coupling regime, the Born's approximation can be employed
\begin{equation}
    \rho_{\text{tot}}(\tau)\approx \rho(\tau)\otimes \rho_{\Phi},
\end{equation}
where $\rho(\tau)=\mathrm{Tr}_{\mathrm{B}}\left[\rho_{\text{tot}}\right]$ is the reduced density matrix of the detector, while $\rho_{\Phi}$ is the state of background fields. In order to simplify the treatment and write down the explicit expressions for the time-evolved state of a UDW detector, we further assume that $[H_{\Phi},\rho_{\Phi}]=0$ as well as $\langle\phi_{\alpha}(x)\rangle=0$, thus the correlation functions of $\phi_{\alpha}$ takes the form
\begin{equation}
\begin{aligned}
    \left\langle\phi_{\alpha}(x(\tau))\phi_{\beta}(x(s))\right\rangle=\delta_{\alpha\beta}\Delta^{+}(\tau-s),
\end{aligned}
\end{equation}
where $\Delta^{+}(\tau-s)=\langle 0| {\phi}[x(\tau)] {\phi}[x(s)]|0\rangle$ is standard Wightman function for a single field, and $\Delta^{-}(\tau)\equiv\Delta^{+}(\tau)^*$ is also defined for later use.
%and the coefficients $c_{i}^{\alpha}$ in \eqref{HI} satisfies the condition 
%\begin{equation}
%    c_{i}^{\alpha}c_{j}^{\beta}\delta_{\alpha\beta}=\delta_{ij}.
%\end{equation}

Under these assumptions, the evolution of the detector density matrix $\rho(\tau)$ is reduced to an integro-differential master equation \cite{Open1,dS15-2}
\begin{equation}
\label{REDFIELD}
    \dot{\rho}(\tau)=g^{2}\sum_{i=1}^3\int_{0}^{\tau}ds\,[\sigma^{i}(s)\rho(s),\sigma^{i}(\tau)]\Delta^{+}(\tau-s)+h.c.
\end{equation}
{One may note that to evaluate the detector state $\rho$ at a specific time $\tau$, we need to know the complete evolution history of $\rho(s)$ from 0 to $\tau$, which reflects the memory effect. By introducing the Markov approximation, which states that the background field has a short correlation time compared to the typical evolution timescale of the detector, one can replace $\rho(s)$ with $\rho(\tau)$ in the integral and obtain a time-local (i.e., memoryless) Redfield equation
\be
\dot{\rho}(\tau)=g^2 \sum_{i=1}^3 \int_0^\tau d s\left[\sigma^i(s) \rho(\tau), \sigma^i(\tau)\right] \Delta^{+}(\tau-s)+\text { h.c. }\no
\ee
Unfortunately, the Redfield equation is unreliable as it fails to maintain the CP (completely positive) property of quantum dynamics, risking a non-positive density matrix \cite{Open1}. To restore CP in Markovian evolution, the RWA approximation is necessary, which eventually recasts the Redfield equation to the so-called Gorini-Kossakowski-Sudarshan-Lindblad (GKSL) form of quantum Markovian master equation of UDW detector \cite{Open1-1}, which is widely used in previous works.

However, many recent works indicate that the RWA approximation often cannot be applied to UDW detectors \cite{Open1-4,Open1-5,dS15-1}. Therefore, it is more appropriate for us to take \eqref{REDFIELD}, where neither the Markov nor the RWA approximation has been employed, as a proper starting point for describing the open dynamics of the UDW detector in a general spacetime background.} The evolution equation (\ref{REDFIELD}) is indeed non-Markovian, in the sense that $\dot{\rho}(\tau)$ depends on the history $\rho(s)$ for all $s\leq\tau$, instead of depending on $\rho(\tau)$ alone\footnote{{We acknowledge that without quantitative measures of the degree, referring to non-Markovianity simply as the existence of a memory effect is vague. Nevertheless, since our ultimate concern is to use quantum Fisher information to estimate the Hubble parameter, we tolerate this vagueness but refer to the excellent review \cite{R-1} for a more precise definition and various measures of the degree of non-Markovianity.}}. Moreover, it is worth emphasizing that the non-Markovian master equation (\ref{REDFIELD}) is accurate to $O(g^{4})$ order \cite{Master1}, guaranteed in the limitation of Born's approximation.

Before proceeding, we employ an alternative basis to simplify the master equation. Introducing three complex vectors $\vec{\xi}^{(\mu)}$ with {$\mu=0,\pm 1$}, so that
\begin{equation} 
\vec{\xi}^{(0)}\equiv\vec{n},\quad \vec{\xi}^{(\mu)}\cdot\vec{\xi}_{(\nu)}^{*}=\delta^{\mu}_{\nu},\quad \vec{\xi}^{(\mu)}\times\vec{\xi}^{(\nu)}=i\sum_{\gamma=0,\pm}\epsilon^{\mu\nu\gamma}\;\vec{\xi}_{(\gamma)}^*,
\label{n}
\end{equation}
where $\vec{\xi}_{(\mu)}^*=\vec{\xi}^{(-\mu)}$, and $\epsilon^{\alpha\beta\gamma}$ is completely anti-symmetric with $\epsilon^{1,0,-1}=1$. It is straightforward to verify that these vectors form a complete basis
\be
\sum_{\mu=0,\pm}(\vec{\xi}^*_{(\mu)})^i(\vec{\xi}^{(\mu)})_j=\delta^{ij},
\ee
by which the original Pauli matrices $\sigma^i$ can be projected to $\sigma^{(\mu)}\equiv\vec{\xi}^{(\mu)}\cdot\vec{\sigma}$, evolving under time translation operator as 
\begin{equation}
   \sigma^{(\mu)}(\tau)= \mathcal{U}_{\tau}[\sigma^{(\mu)}]=e^{iH_{S}\tau}\left(\vec{\xi}^{(\mu)}\cdot\vec{\sigma}\right) e^{-iH_{S}\tau}=e^{i\mu\omega \tau}\sigma^{(\mu)}.
\end{equation}
In terms of $\sigma^{(\mu)}$, the non-Markovian master equation \eqref{REDFIELD} can now be recasted into 
\begin{equation}
\label{REDFIELD2}
    \dot{\rho}(\tau)=g^{2}\sum_{\mu=0,\pm}\int_{0}^{\tau}ds\,[\sigma^{\dag}_{(\mu)}\rho(s),\sigma^{(\mu)}]\Delta^{+}(\tau-s)e^{i\mu\omega(\tau-s)}+h.c.
\end{equation}

For a single detector (qubit), its density matrix can be decomposed in a Bloch-like form
\begin{equation}
    \rho(\tau)=\frac{1}{2}\left(1+\sum_{\mu=0,\pm}v_{(\mu)}(\tau)\sigma^{(\mu)}\right).
    \label{rho}
\end{equation}
Benefited from the fact that $\sigma^{(\mu)}$ are eigenvectors of $\mathcal{U}_{\tau}$ which also commutes with \eqref{REDFIELD}, the time evolution of coefficients $v_{(\mu)}(\tau)$ in new basis now decouples \cite{Open1-1}. For instance, without loss of generality, we can choose $\vec{n}=\hat{z}$ in \eqref{d-Hamiltonian}, which gives $\vec{\xi}^{(0)}=\hat{z}$ and $\vec{\xi}^{(\pm)}=\frac{1}{\sqrt{2}}(\hat{x}\pm i\hat{y})$. Then the decomposition \eqref{rho} is explicitly given as
\begin{equation}
    \rho=\frac{1}{2}\begin{bmatrix}
    1+v_{(0)} && \sqrt{2}v_{(+)}
    \\
    \sqrt{2}v_{(-)} && 1-v_{(0)}
    \end{bmatrix},
    \label{normalrho}
\end{equation}
where $v_{(0)}=\rho_{11}-\rho_{00}$ determines the time-evolution of diagonal elements, and $v_{(\pm)}$ determines the time-evolution of off-diagonal elements of detector qubit.

We are now in a position to explicitly solve the non-Markovian master {equation} \eqref{REDFIELD2}. Noting that  \eqref{REDFIELD2} admits a form of convolution, we employ a Laplace transformation and obtain
\begin{equation}
    z\tilde{\rho}(z)-\rho(0)=g^{2}\sum_{\mu=0,\pm}[\sigma^{\dag}_{(\mu)}\tilde{\rho}(z),\sigma^{(\mu)}]\tilde{\Delta}^{+}(z-i\mu\omega)+h.c.
    \label{REDFIELD3}
\end{equation}
where $\tilde{\Delta}^{+}$ is the Laplace-transformed Wightman function, and the Laplace-transformed density matrix $\tilde{\rho}(z)$ from \eqref{normalrho} admits a similar Bloch-decomposition
\begin{equation}
    \tilde{\rho}(z)=\frac{1}{2}\left(\frac{1}{z}+\sum_{\mu=0,\pm}\tilde{v}_{(\mu)}(z)\sigma^{(\mu)}\right).
\end{equation}

The time-evolution of the coefficients $\tilde{v}_{(\mu)}$ can be given explicitly (for details see Appendix \ref{appendixA})
\begin{equation}
  \left\{
    \begin{aligned}
        \tilde{v}_{(0)}(z)&=\frac{\left.v_{(0)}\right|_{\tau=0}+2g^{2}M(z)/z}{z+2g^{2}N_{(0)}(z)},\\
        \tilde{v}_{(\pm)}(z)&=\frac{\left.v_{(\pm)}\right|_{\tau=0}}{z+2g^{2}N_{(\pm)}(z)},
        \label{v}
    \end{aligned}
    \right.
\end{equation}
where the functions $M(z)$, $N_{(\mu)}(z)$ are defined by
\begin{equation}
\begin{aligned}
    M(z)\equiv&\tilde{\Delta}^{+}(z+i\omega)-\tilde{\Delta}^{-}(z+i\omega)+\tilde{\Delta}^{-}(z-i\omega)-\tilde{\Delta}^{+}(z-i\omega),\\
     N_{(0)}(z)\equiv&\tilde{\Delta}^{+}(z+i\omega)+\tilde{\Delta}^{-}(z+i\omega)+\tilde{\Delta}^{+}(z-i\omega)+\tilde{\Delta}^{-}(z-i\omega),\\
     N_{(\pm)}(z)\equiv&\tilde{\Delta}^{+}(z\mp i\omega)+\tilde{\Delta}^{-}(z\mp i\omega)+\tilde{\Delta}^{+}(z)+\tilde{\Delta}^{-}(z).
     \label{MN}
\end{aligned}
\end{equation}
Except been written in a more efficient basis, we can check that the results \eqref{v} are consistent with those were given in \cite{dS15-2}.%We may use $p^{\pm}$ to denote $p\pm i\omega$ in the following discussion due to their frequent appearance. 

Taking inverse Laplace transformation on $\tilde{v}_{(\mu)}(z)$, we recover the coefficients $v_{(\mu)}$ as
\begin{equation}
    v_{(\mu)}(\tau)=\frac{1}{2\pi i}\int_{c-i\infty}^{c+i\infty}\mathrm{d}p\,\tilde{v}_{(\mu)}(z)e^{z\tau},
    \label{InverseLaplaceTransformation}
\end{equation}
where $c$ is a real constant larger than the real part of all the singularities of the integrand \cite{Toolbox}. Obviously, the time-evolving coefficients $v_{(\mu)}(\tau)$ should depend on the {specific} correlation functions $\Delta^{\pm}(\tau)$.

\subsection{Correlation functions of scalar fields within $\alpha$-vacua}
\label{sec:corrlation}

We consider the comoving UDW detector (qubit) interacting with a massless scalar field in expanding de Sitter space\footnote{Strictly speaking, no de Sitter-invariant vacuum exists for a massless scalar field. Nevertheless, we can evade this subtlety by assuming a tiny but nonzero mass, with a price of failed Markovian limit that should be paid, as revealed by \cite{dS13}.}. The time evolution is investigated along the direction of planar coordinates, in terms of which the spacetime line element is 
\be
{d} s^2=-\mathrm{d} \tau^2+e^{2 H \tau} \mathrm{~d} \mathbf{x}^2=\frac{1}{H^2 \eta^2}\left(-\mathrm{d} \eta^2+\mathrm{d} \mathbf{x}^2\right),
\ee
where $H>0$ is the Hubble parameter, and the conformal time $\eta$ related to comoving time $\tau$ by $\eta=- e^{-H \tau}/H$, which represents a flat slicing of spacetime.

As explained in the introduction, we aim to resolve the time evolution of the UDW detector respecting the de Sitter-invariant vacua of a real scalar field. 
{The most general class is the so-called $\alpha$-vacua \cite{dS3}, which is a family of CPT-invariant vacuum states $|\alpha\ra$ labeled by the single parameter $\alpha<0$. 

The physical significance of $\alpha$-vacua primarily stems from their key role in constructing a holographic understanding of quantum gravity in de Sitter space. While the unique Bunch-Davies vacuum $|\text{BD}\ra\equiv|\alpha\rightarrow-\infty\ra$ extrapolates to the standard Minkowski vacuum in the limit of vanishing cosmological constant, the general vacua $|\alpha\rangle$ are realized as squeezed states over $|\text{BD}\ra$. Within the context of dS/CFT correspondence \cite{dS15}, one can extract CFT correlation functions from the $n$-point correlator in de Sitter. The change of the vacuum $|\alpha\ra$ in the bulk of de Sitter then corresponds to a marginal deformation of the associated CFT \cite{dS10}. From the holographic view \cite{dS16}, such rich vacuum choices bring further interpretation to the celebrated Ryu-Takayanagi scheme \cite{dS17} of entanglement entropy in quantum gravity.

An even stronger motivation to study $\alpha$-vacua arises from the trans-Planckian physics in inflationary cosmology \cite{dS18}. Conventionally, quantum fluctuations are believed to originate in the infinite past (at conformal time $\eta\rightarrow-\infty$) with an infinitely short wavelength, which indicates the Bunch-Davies vacuum as an inflaton initial condition. However, in the real world, this cannot be true as fundamentally new physics is expected at the Planck scale $\Lambda$. Thus, the modified initial condition must be imposed at a short-distance cut-off at the Planck scale \cite{alpha-0} (e.g., a proper time $\eta_0\sim-{\Lambda}/{H}$), manifested by the alternative non-Bunch-Davies states. In particular, for their naturally respecting spacetime isometry, de Sitter $\alpha$-vacua become an attractive possibility \cite{dS8,dS8+,dS9,alpha-1,alpha-2,alpha-3}, especially those with $\alpha$ constrained by the fundamental scale like $e^\alpha \sim {H}/{\Lambda}$. Moreover, through the inflation stretching, the trans-Planckian frequencies resulting from alternative vacuum choices are expected to redshift down and leave an observable imprint (e.g., the non-Gaussianity deviation of CMB in Planck satellite experiments \cite{alpha-4}), providing an important window to complete the theory of quantum gravity in de Sitter space.}

Within these $\alpha$-vacua, the scalar Wightman function $\Delta^{+}(x,y;\alpha)=\langle \alpha| {\phi}(x) {\phi}(y)|\alpha\rangle$ are related to the ones in Bunch-Davies vacuum $\Delta^{+}(x,y)=\langle \text{BD}|{\phi}(x) {\phi}(y)|\text{BD}\rangle$ like \cite{dS8,dS10,dS12}: 
\begin{equation}
    \Delta^{+}(x,y;\alpha)=\frac{1}{1-e^{2\alpha}}\left[\Delta^{+}(x,y)+e^{2\alpha}\Delta^{+}(y,x)+e^{\alpha}(\Delta^{+}(x,y_A)+\Delta^{+}(x_A,y))\right],
\end{equation}
where $x_A$ is the antipodal point of $x$, and the parameter $\alpha$ in $\Delta^{+}(x,y;\alpha)$ labels Wightman functions with respect to different $\alpha-$vacua choice. As $\alpha\rightarrow{-\infty}$, the vacua $|\alpha\ra$ approaches to Bunch-Davies one $|\text{BD}\ra$, thus leading $\Delta^{+}(x,y;\alpha)$ reduce to $\Delta^{+}(x,y)$. 

For a comoving detector parameterized by its own proper time, we have Wightman functions in Bunch-Davies vacuum \cite{UDW3}
\begin{equation}
\left\{    \begin{aligned}
        \Delta^{+}(x(\tau),x(s))=-\frac{H^2}{16\pi^{2}\sinh^{2}\left(\frac{H(\tau-s)-i\epsilon}{2}\right)},\\
        \Delta^{+}(x_{A}(\tau),x(s))=\frac{H^2}{16\pi^{2}\cosh^{2}\left(\frac{H(\tau-s)-i\epsilon}{2}\right)},\\
        \Delta^{+}(x(\tau),x_{A}(s))=\frac{H^2}{16\pi^{2}\cosh^{2}\left(\frac{H(\tau-s)+i\epsilon}{2}\right)},
    \end{aligned}
    \right.
\end{equation}
where the parameter $\epsilon$ regularize the divergence at $\tau\rightarrow 0$ with a cutoff scale $\epsilon'= \epsilon H$. 

Applying a Laplace transformation on $\Delta^{+}(x(\tau),x(0);\alpha)$, we obtain
\begin{equation}
    \tilde{\Delta}^{\pm}(z;\alpha)=\frac{1}{1-e^{2\alpha}}\left[E_{(\pm)}(z)+e^{2\alpha}E_{(\mp)}(z)+2e^{\alpha}F(z)\right]
\label{core},
\end{equation}
where
\begin{equation}
    \begin{aligned}
F(z)\equiv&\int_{0}^{\infty}\mathrm{d}\tau\,\left(\frac{\Delta^{+}(x_{A}(\tau),x(0))+\Delta^{+}(x(\tau),x_{A}(0))}{2}\right)e^{-z\tau},\\
E_{(\pm)}(z)\equiv&\int_{0}^{\infty}\mathrm{d}\tau\,\Delta^{+}(x(\pm \tau),x(0))e^{-z\tau},
    \end{aligned}
\end{equation}
which can be explicitly integrated out as (for details see Appendix \ref{appendixB})
\begin{equation}
    \begin{aligned}
        &E_{(\pm)}(z)=\frac{H}{8\pi^2}\left(\pm\coth{\frac{i\epsilon}{2}}+2z\sum_{n=0}^{\infty}\frac{e^{\pm in\epsilon}}{z+nH}-1\right),\\
        &F(z)=\frac{H}{8\pi^2}\left(2z\sum_{n=0}^{\infty}\frac{(-1)^{n}\cos{n\epsilon}}{z+nH}-1\right).
        \label{EF}
    \end{aligned}
\end{equation}
Substituting these back into \eqref{core}, we obtain the Laplace-transformed correlation functions $\tilde{\Delta}^{\pm}(z;\alpha)$ as
\begin{equation}
    \begin{aligned}
        \tilde{\Delta}^{\pm}(z;\alpha)=\frac{H}{8\pi^2}\left(-\sum_{n=0}^{\infty}\frac{[\coth{\alpha}+(-1)^{n}\csch{\alpha}]\cos{n\epsilon}\mp i\sin{n\epsilon}}{(1+nH/z)/2}-\frac{(1+e^{\alpha})^2}{1-e^{2\alpha}}\pm\coth{\frac{i\epsilon}{2}}\right).
    \end{aligned}
    \label{core2}
\end{equation}
We observe that the functions have infinite simple poles at $z_n=-nH$ for $n\in\mathbb{Z}^{+}$, with residues like
\begin{equation}
    \begin{aligned}
       \mathfrak{Res}\left[\tilde{\Delta}^{\pm}(z;\alpha),z_n\right]=\frac{H^2}{4\pi}\left(n\left[\coth{\alpha}+(-1)^{n}\csch{\alpha}\right]\cos{n\epsilon}\mp in\sin{n\epsilon}\right).
       \label{res1}
    \end{aligned}
\end{equation}

With the help from \eqref{res1}, we can determine the pole structures of the functions $M(z;\alpha)$ and $N_{(\mu)}(z;\alpha)$ respecting different $\alpha-$vacua. For instance, we start with $N_{(0)}(z;\alpha)$ defined by \eqref{MN} as
\begin{equation}
    \begin{aligned}
        N_{(0)}(z;\alpha)\equiv\tilde{\Delta}^{+}(z+i\omega;\alpha)+\tilde{\Delta}^{-}(z+i\omega;\alpha)+\tilde{\Delta}^{-}(z-i\omega;\alpha)+\tilde{\Delta}^{+}(z-i\omega;\alpha)
    \end{aligned}.
\end{equation}
They have infinite simple poles located at $z_n^{\pm}=-Hn\pm i\omega$ for $n\in \mathbb{Z}^{+}$, with corresponding residues
\begin{equation}
    \begin{aligned}
          \mathfrak{Res}\left[N_{(0)}(z;\alpha),z_n^{\pm}\right]=\frac{H^2n[\coth{\alpha}+(-1)^{n}\csch{\alpha}]\cos{n\epsilon}}{2\pi^2}
    \end{aligned}.
\end{equation}
By looking at the functions
\begin{equation}
    M(z;\alpha)=\tilde{\Delta}^{+}(z+i\omega;\alpha)-\tilde{\Delta}^{-}(z+i\omega;\alpha)+\tilde{\Delta}^{-}(z-i\omega;\alpha)-\tilde{\Delta}^{+}(z-i\omega;\alpha),
\end{equation}
one can tell that they have poles located at $z_n^\pm$, whose residues can be given as
\begin{equation}
    \begin{aligned}
     \mathfrak{Res}\left[M(z;\alpha),z_n^\pm\right]=\pm\frac{iH^2n\sin{n\epsilon}}{2\pi^2}.
    \end{aligned}
\end{equation}
The functions $N_{(\pm)}(z;\alpha)$ have poles at both $z_n$ and $z_n^\pm$, with residues:
\begin{equation}
    \begin{aligned}
         \mathfrak{Res}\left[N_{(\pm)}(z;\alpha),z_n\right]=\mathfrak{Res}\left[N_{(\pm)}(z;\alpha),z_n^\pm\right]=\frac{H^2n[\coth{\alpha}+(-1)^{n}\csch{\alpha}]\cos{n\epsilon}}{2\pi^2}.
    \end{aligned}
\end{equation}

Collecting the above results, we eventually obtain
\begin{equation}
\left\{
    \begin{aligned}
    M(z;\alpha)&=\sum_{n=1}^{\infty}\mathfrak{M}_{n}\left(\frac{1}{z+nH-i\omega}-\frac{1}{z+nH+i\omega}\right)+\text{regular},\\
N_{(0)}(z;\alpha)&=\sum_{n=1}^{\infty}\mathfrak{N}_{n}\left(\frac{1}{z+nH-i\omega}+\frac{1}{z+nH+i\omega}\right)+\text{regular},\\
N_{(\pm)}(z;\alpha)&=\sum_{n=1}^{\infty}\mathfrak{N}_{n}\left(\frac{1}{z+nH}+\frac{1}{z+nH\mp i\omega}\right)+\text{regular},
    \end{aligned}
    \label{poleMN}
    \right.
\end{equation}
where the 'regular' parts are holomorphic in $\mathbb{C}$, and two coefficient functions
\begin{equation}
    \mathfrak{M}_{n}=\frac{iH^2n \sin{n\epsilon}}{2\pi^2},~~~~~~~ \mathfrak{N}_{n}=\frac{H^2n \cos{n\epsilon}\left[(-1)^{n}\csch{\alpha}+\cosh{\alpha}\right]}{2\pi^2\sinh{\alpha}}
    \label{rR}
\end{equation}
are introduced to simplify the expression.

\subsection{Dynamics of a UDW detector in de Sitter Space}
\label{Solving the Dynamics of the Qubit}

To resolve the non-Markovian dynamics of the comoving UDW detector, we need to substitute \eqref{poleMN} into \eqref{v} and then evaluate the inverse Laplace transformation in \eqref{InverseLaplaceTransformation}. Here, the coefficient functions $v_{(\mu)}(\tau;\alpha)$ provide the time evolution of the detector's density matrix through \eqref{rho}. Determined by the pole structure of $M(z;\alpha)$ and $N_{(\mu)}(z;\alpha)$ (with $\mu=0,\pm$), we can recognize that the singularities of $\tilde{v}_{(\mu)}(z;\alpha)$ are discrete, allowing us to express \eqref{InverseLaplaceTransformation} as a sum of residues:
\begin{equation}
    v_{(\mu)}(\tau;\alpha)=\frac{1}{2\pi i}\int_{c-i\infty}^{c+i\infty}\mathrm{d}z\,\tilde{v}_{(\mu)}(z;\alpha)e^{z\tau}=\sum_{k}\mathfrak{Res}\left[\tilde{v}_{(\mu)}(z;\alpha)e^{z\tau},z_{(\mu),k}\right],
    \label{InverseLaplacetransformation2}
\end{equation}
where $\{z_{(\mu),k}\}_{k=1}^{\infty}$ denote the poles of functions $\tilde{v}_{(\mu)}(z;\alpha)$. As shown in \eqref{v} that the evolution of coefficients $v_{(\mu)}$ decouples, we treat the diagonal part $v_{(0)}$ and the off-diagonal part $v_{(\pm)}$ (e.g., of the detector density matrix \eqref{normalrho}) separately.

\subsubsection{The diagonal part}
\label{sec:Diagonal}
We first determine the pole structure of $\tilde{v}_{(0)}(z;\alpha)$, {which is now given as}
\begin{equation}
   \tilde{v}_{(0)}(z;\alpha)=\frac{\left.v_{(0)}\right|_{\tau=0}+2g^{2}M(z;\alpha)/z}{z+2g^{2}N_{(0)}(z;\alpha)}.
    \label{v0}
\end{equation}

\begin{figure}[htbp]
\centering
\includegraphics[width=.55\textwidth]{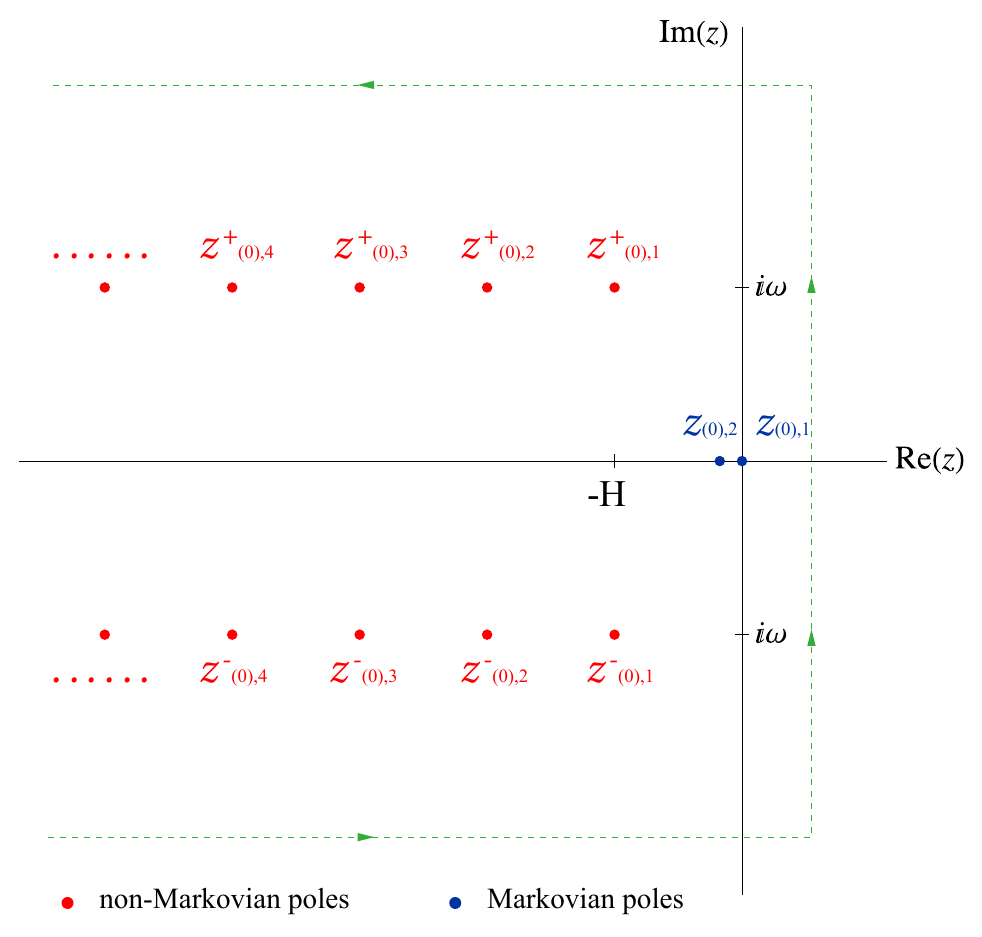}
\caption{The pole structure of the Laplace-transformed diagonal coefficient $\tilde{v}_{(0)}(z;\alpha)$, where two blue points represent Markovian poles $z_{(0),1}$ and $z_{(0),2}$, and two infinite sets of red points label the non-Markovian poles $z_{(0),n}^+$ and $z_{(0),n}^-$ ($n\in\mathbb{Z}^{+}$).}
\label{diagonalpoles}
\end{figure}

\textbf{Markovian poles}---Since $M(z;\alpha)$ and $N_{(0)}(z;\alpha)$ are analytic at $z=0$, we can find  two poles of the function $\tilde{v}_{(0)}(z;\alpha)$ located at $z_{(0),1}=0$ and $z_{(0),2}$, where the later one is determined by the equation 
\begin{equation}
    z_{(0),2}+2g^{2}N_{(0)}\left(z_{(0),2};\alpha\right)=0
    \label{pm}
\end{equation}
around the point $z=0$. We call $z_{(0),1}=0$ and $z_{(0),2}$ Markovian poles as depicted in Fig.\ref{diagonalpoles}. 

The residue at $z_{(0),1}=0$ is easy to determine as
\begin{equation}
    \mathfrak{Res}\left[\tilde{v}_{(0)}(z;\alpha),z_{(0),1}\right]=\frac{M(0;\alpha)}{N_{(0)}(0;\alpha)}.
\end{equation}
The residue at $z_{(0),2}$ is more subtle. Based on complex analysis, the result {is}
\begin{equation}
    \mathfrak{Res}\left[\tilde{v}_{(0)}(z;\alpha),z_{(0),2}\right]=\frac{v_{(0)}|_{\tau=0}+2g^{2}M\left(z_{(0),2};\alpha\right)/z_{(0),2}}{1+2g^{2}N'_{(0)}\left(z_{(0),2};\alpha\right)}.
    \label{respm}
\end{equation}
where {``$\,^\prime\,$''} indicates a derivative with respect to complex variable $z$. Since the the numerator contains contributions of the form $g^{2}/z_{(0),2}$, we have to expand $z_{(0),2}$ to $\mathcal{O}(g^{6})$ to guarantee that the result is accurate to $\mathcal{O}(g^{4})$. From \eqref{pm}, we have the expansion 
\begin{equation}
    z_{(0),2}\approx-2g^{2}N_{(0)}(0;\alpha)\left(1-2g^{2}N'_{(0)}(0;\alpha)\right)+\mathcal{O}(g^6).
\end{equation}
Combining with \eqref{respm}, we obtain
\begin{equation}
    \mathfrak{Res}\left[\tilde{v}_{(0)}(z;\alpha),z_{(0),2}\right]=[1-2g^{2}N'_{(0)}(0;\alpha)]v_{(0)}(0;\alpha)+2g^{2}M'(0;\alpha)-\frac{M(0;\alpha)}{N_{(0)}(0;\alpha)}+\mathcal{O}(g^{4}),
    \label{respm2}
\end{equation}
where $v_{(0)}(0;\alpha)=v_{(0)}|_{\tau=0}$ represents the initial value of the coefficient. It is important to note that our evaluation \eqref{respm2} {differs} from those in \cite{dS15-2}, where the second and third terms are absent for an accelerating UDW qubit. {However, these additional terms are necessary to match the detector state $\rho(\tau)$ with the initial condition as $\tau\rightarrow 0^{+}$} in the de Sitter scenario. 

\textbf{Non-Markovian poles}---Since $g$ is small, the only remaining place for $\tilde{v}_{(0)}(z;\alpha)$ to be singular is the neighborhood of $z_{n}^\pm=-nH\pm i\omega$, where $M(z;\alpha)$ and $N_{(0)}(z;\alpha)$ diverge. Therefore, we anticipate that additional sets of (infinite) poles $z_{(0),n}^\pm$ close to $z_{n}^\pm$ should exist, which we identify as non-Markovian poles based on insights from \cite{dS15-2}, suggesting their contributions can be neglected in the Markov-approximated master equation for an accelerating UDW qubit. 

From \eqref{poleMN}, we know that in a neighborhood of $z_{n}^{\pm}$,
\begin{equation}
    M(z;\alpha) = \pm\frac{\mathfrak{M}_{n}}{z-z_{n}^{\pm}}+f_M(z),~~~~ N_{(0)}(z;\alpha) = \frac{\mathfrak{N}_n}{z-z_{n}^{\pm}}+f_{N_{(0)}}(z),
    \label{splitMN}
\end{equation}
where  $f_{M}(z)$ and $f_{N_{(0)}}(z)$ are analytic near $z^{\pm}_{n}$, whose details need not be worried about as they only generate $\mathcal{O}(g^4)$ contributions in the final results. 

The non-Markovian poles $z_{(0),n}^\pm$ are determined from
\begin{equation}
    z_{(0),n}^{\pm}+2g^{2}N_{(0)}\left(z_{(0),n}^{\pm};\alpha\right)=0,
\end{equation}
which can be resolved by the decomposition \eqref{splitMN}, leading to
\begin{equation}
    z_{(0),n}^{\pm}-z_{n}^{\pm}=-\frac{2g^{2}~\mathfrak{N}_{n}}{z_{n}^{\pm}+2g^{2}~f_{N_{0}}(z_{n}^{\pm})}.
\end{equation}
{Ignoring higher order contributions in $g$, we obtain} (depicted by red points in Fig.\ref{diagonalpoles})
\begin{equation}
    z_{(0),n}^{\pm}=-nH\pm i\omega+\frac{2g^{2}\mathfrak{N}_{n}}{nH\mp i\omega}+\mathcal{O}(g^{4}).
    \label{diagonalnonmarkovpole}
\end{equation}

{Once the non-Markovian poles $z_{(0),n}^{\pm}$ are specified}, the residues of coefficient functions $\tilde{v}_{(0)}(z;\alpha)$ at these poles can be calculated from
\begin{equation}
    \mathfrak{Res}\left[\tilde{v}_{(0)}(z;\alpha),z_{(0),n}^{\pm}\right]=\frac{\left.v_{(0)}\right|_{\tau=0}+2g^{2}M\left(z_{(0),n}^{\pm};\alpha\right)/z_{(0),n}^{\pm}}{1+2g^{2}N'_{(0)}\left(z_{(0),n}^{\pm};\alpha\right)}.
    \label{resnonMv}
\end{equation}
Substituting the straightforwardly calculated results 
\begin{equation}
\begin{aligned}
    2g^2 N_{(0)}'\left(z_{(0),n}^{\pm};\alpha\right)&=-\frac{(nH \mp i\omega)^2}{2g^2 \mathfrak{N}_n}+\mathcal{O}(g^2),\\ 
    2g^2 M\left(z_{(0),n}^{\pm};\alpha\right)/z_{(0),n}^{\pm}&= \mp\frac{\mathfrak{M}_{n}}{\mathfrak{N}_{n}}+\mathcal{O}(g^2)
\end{aligned}
\end{equation}
into \eqref{resnonMv}, we eventually obtain the desired residue of $\tilde{v}_{(0)}(z;\alpha)$ at non-Markovian poles as
\begin{equation}
     \mathfrak{Res}\left[\tilde{v}_{(0)}(z;\alpha),z_{(0),n}^{\pm}\right]=\frac{2g^{2}\left[\pm \mathfrak{M}_{n}-\mathfrak{N}_{n}v_{(0)}(0;\alpha)\right]}{(nH \mp i\omega)^2}+\mathcal{O}(g^{4}).
\end{equation}

In summary, we find that the functions $\tilde{v}_{(0)}(z;\alpha)$ have two Markovian poles located at $z_{(0),1}$ and $z_{(0),2}$, as well as two infinite sequences of non-Markovian poles situated at $z_{(0),n}^+$ and $z_{(0),n}^-$ for $n>0$, respectively. Figure~\ref{diagonalpoles} illustrates the distribution of these poles on the complex plane.

After determining the complete pole structure of the Laplace-transformed diagonal coefficient $\tilde{v}_{(0)}(z;\alpha)$, as well as the corresponding residues, we are positioned to evaluate the inverse Laplace transformation \eqref{InverseLaplacetransformation2}, which explicitly provides the diagonal part of a comoving UDW detector state \eqref{normalrho} as
\begin{equation}
\begin{aligned}
    &v_{(0)}(\tau;\alpha)=\gamma_{(0)}\left(e^{-\Gamma_{(0)}\tau}-1\right)+v_{(0)}(0;\alpha)e^{-\Gamma_{(0)}\tau}
    \\
   & +2g^{2}\left\{\left[M'(0;\alpha)-N'_{(0)}(0;\alpha)v_{(0)}(0;\alpha)\right]e^{-\Gamma_{(0)}\tau}+\mathfrak{Re}\left[\sum_{n=1}^{\infty}\frac{2\left[\mathfrak{M}_{n}-\mathfrak{N}_{n}v_{(0)}(0;\alpha)\right]e^{(i\omega-nH) \tau}}{(nH-i \omega)^2}\right]\right\},
    \label{v0t}
\end{aligned}
\end{equation}
where $\mathcal{O}(g^4)$ contributions have been ignored, $\mathfrak{Re}$ takes the real part of variable, and two important functions $\gamma_{(0)}$ and $\Gamma_{(0)}$ are defined by (for details see Appendix \ref{appdixD})
\begin{equation}
\left\{
    \begin{aligned}
        \gamma_{(0)}&\equiv-\frac{M(0;\alpha)}{N_{(0)}(0;\alpha)}=-\frac{\sinh{\alpha}\sinh{\left(\frac{\pi\omega}{H}\right)}}{1+\cosh{\alpha}\cosh{\left(\frac{\pi\omega}{H}\right)}}+\mathcal{O}(\epsilon\log{\epsilon}),\\
        \Gamma_{(0)}&\equiv2g^{2}N_{(0)}(0;\alpha)=-\frac{g^{2}\omega}{\pi}\left[\frac{\sinh{\alpha}\sinh{\left(\frac{\pi\omega}{H}\right)}}{1+\cosh{\alpha}\cosh{\left(\frac{\pi\omega}{H}\right)}}\right]^{-1}+\mathcal{O}(g^{2}\epsilon\log{\epsilon}).
        \label{coefficient1}
    \end{aligned}
    \right.
\end{equation}
Be aware that the first line of \eqref{v0t} is attributed to the contribution of the residues of Markovian poles; thus, it solves the Markov-approximated evolution equation, while the second line of \eqref{v0t} represents the non-Markovian correction. 

Before proceeding further, we would like to revisit the necessity of additional terms in \eqref{respm2} by demonstrating that the condition
\begin{equation}
    \lim_{\tau\rightarrow 0^{+}}v_{(0)}(\tau)=v_{(0)}(0),
    \label{initialconsistency}
\end{equation}
is now fulfilled by \eqref{v0t}. Noting that the Markovian contribution (i.e. the first line of \eqref{v0t}) automatically satisfies the condition \eqref{initialconsistency}, we thus need to prove the following identities at $\tau=0$
\begin{equation}
    M'(0;\alpha)-N'_{(0)}(0;\alpha)v_{(0)}(0;\alpha)+2\;\mathfrak{Re}\left[\sum_{n=1}^{\infty}\frac{\mathfrak{M}_{n}-\mathfrak{N}_{n}v_{(0)}(0;\alpha)}{(nH-i \omega)^2}\right]=0.
    \label{initialconsistency2}
\end{equation}
Constructing the integrals:
\begin{equation}
    \chi_{M}(r)\equiv\frac{1}{2\pi i}\oint_{C_{r}}\frac{M(z;\alpha)}{z^{2}}\,\mathrm{d}z, ~~~~~~~~ \chi_{N_{(0)}}(r)\equiv\frac{1}{2\pi i}\oint_{C_{r}}\frac{N_{(0)}(z;\alpha)}{z^{2}}\,\mathrm{d}z,
\end{equation}
where $C_{r}$ is the circle centered at $z=0$ with radius $r$. Based on the growth behavior of $M(z;\alpha)$ and $N_{(0)}(z;\alpha)$ as $|z|\rightarrow \infty$, one can check that
\begin{equation}
    \lim_{r\rightarrow \infty}\chi_{M}(r)=\lim_{r\rightarrow \infty}\chi_{N_{(0)}}(r)=0.
\end{equation}
Using the residue theorem, we have
\begin{equation}
\begin{aligned}
     M'(0;\alpha)+\mathfrak{Re}\left[\sum_{n=1}^{\infty}\frac{2\; \mathfrak{M}_{n}}{(nH-i\omega)^2}\right]=0,~~~~~
     N'_{(0)}(0;\alpha)+\mathfrak{Re}\left[\sum_{n=1}^{\infty}\frac{2\; \mathfrak{N}_{n}}{(nH-i\omega)^2}\right]=0,
\end{aligned}
\end{equation}
respectively, which completes the proof.

By the end of this section, we would like to show that \eqref{v0t} can be further simplified by noting the fact that the inhomogeneous contributions from $\mathfrak{M}_{n}$ and $M'(0;\alpha)$ are negligible. To see this, we separate out the contribution of $\mathfrak{M}_{n}$ from \eqref{v0t}, which can be attributed to a {function}
\begin{equation}
\mathfrak{E}(\tau)\equiv\mathfrak{Re}\left[\sum_{n>0}\frac{4\;\mathfrak{M}_{n}e^{(i\omega-nH)\tau}}{(nH-i\omega)^{2}}\right]=\mathfrak{Re}\left[\sum_{n>0}\frac{2 i H^2 n\sin{n\epsilon}}{\pi^{2}(nH-i\omega)^{2}}e^{(i\omega-nH)\tau}\right].
\label{f}
\end{equation}
{We refer to $\mathfrak{E}(\tau)$ as an "estimator" since its value bounds both $\mathfrak{M}_{n}$ and $M'(0;\alpha)$.} {Since $\sin{n\epsilon}\sim \mathcal{O}(\epsilon)$ as $\epsilon\rightarrow 0$, it's not hard to prove that for $\tau\in[0,\infty)$, we have $\mathfrak{E}(\tau)\sim \mathcal{O}(\epsilon\log{\epsilon})$} (for details see Appendix \ref{sec:estimation}). Since we have $M'(0;\alpha)=-\mathfrak{E}(0)/2 \sim \mathcal{O}(\epsilon\log{\epsilon})$ as a special case, we justify our claim.

Based on the above analysis, we introduce the non-Markovian contribution functions $S_{(0)}(\tau;\alpha)$ as
\begin{equation}
    S_{(0)}(\tau;\alpha)\equiv\mathfrak{Re}\left[\sum_{n=1}^{\infty}\frac{4\;\mathfrak{N}_{n}e^{(i\omega-nH)\tau}}{(nH-i\omega)^{2}}\right],
    \label{s0}
\end{equation}
which allow us to write $v_{(0)}(\tau;\alpha)$ \eqref{v0t} in a concise way
\bea
    v_{(0)}(\tau;\alpha)&\equiv&v_{(0)}^{\text{Markovian}}+v_{(0)}^{\text{non-Markovian}}\no\\
    &=&\gamma_{(0)}\left(e^{-\Gamma_{(0)}\tau}-1\right)+v_{(0)}(0;\alpha)e^{-\Gamma_{(0)}\tau}\no\\
  &&~~~~~~~~~~~~~~~~~~ -g^{2}\bigg[S_{(0)}(0;\alpha)e^{-\Gamma_{(0)}\tau}-S_{(0)}(\tau;\alpha)\bigg]v_{(0)}(0;\alpha),
    \label{v0t2}
\eea
whereby comparing with \eqref{v0t}, we know the first line still respects the Markov-approximated master equation, i.e., the time-evolution governed by the diagonal-decay-rate $\Gamma_{(0)}$, while the functions $S_{(0)}(\tau;\alpha)$ alone determine all non-Markovian contributions. 

Within de Sitter context, the non-Markovian contribution functions $S_{(0)}(\tau;\alpha)$ can be given explicitly as
\bea
    S_{(0)}(\tau;\alpha)&=&-\mathfrak{Re}\left\{\frac{e^{-i\omega \tau}}{\pi^2\sinh{\alpha}}\left[2\phi\left(-e^{-H\tau},\frac{i\omega}{H}\right)\right.\right.\no\\
    &&~~~~~~~~~~~~~~\left.\left.+\cosh{\alpha}\left(\phi\left(e^{-H\tau-i\epsilon},\frac{i\omega}{H}\right)+\phi\left(e^{-H\tau+i\epsilon},\frac{i\omega}{H}\right)\right) \right]\right\},
\eea
where
\begin{equation}
    \phi(z,x)\equiv\sum_{n=1}^{\infty}\frac{nz^{n}}{(x+n)^{2}}=\Phi(z,1,x)-x\,\Phi(z,2,x).
\end{equation}
with $\Phi$ is the Lerch's Transcendent \cite{NIST}.

\subsubsection{The off-diagonal part}
\begin{figure}[htbp]
\centering
\includegraphics[width=.55\textwidth]{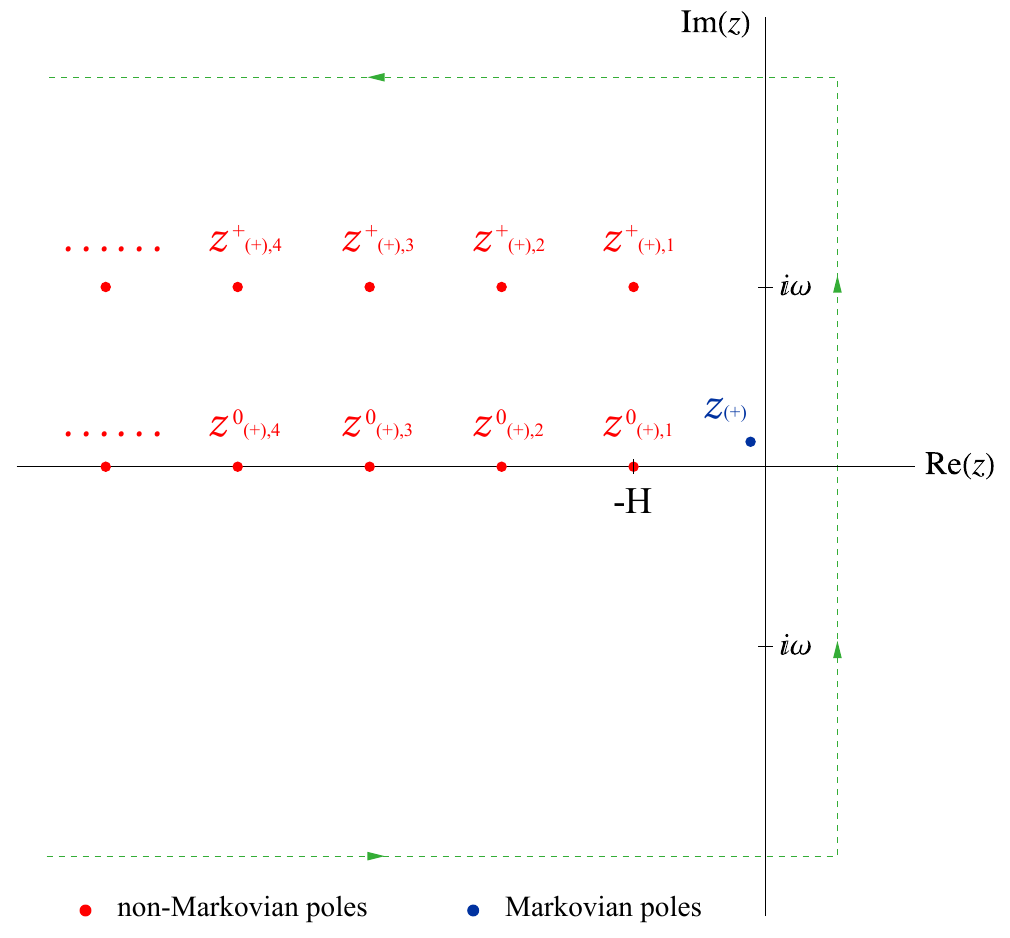}
\caption{The pole structure of the Laplace-transformed non-diagonal coefficient $\tilde{v}_{(+)}(z;\alpha)$, where one blue point represents Markovian poles $z_{(+)}$, and two infinite sets of red points label the non-Markovian poles $z_{(+),n}^0$ and $z_{(+),n}^+$ ($n\in\mathbb{Z}^{+}$).}
\label{offdiagonalpoles}
\end{figure}

From the definition, we know that $v_{(\pm)}(\tau)^*=v_{(\mp)}(\tau)$. Therefore, it is sufficient for us to study the poles structure of $\tilde{v}_{(+)}(z;\alpha)$, which can be read from \eqref{v} as
\begin{equation}
    \tilde{v}_{(+)}(z;\alpha)=\frac{\left.v_{(+)}\right|_{\tau=0}}{z+2g^{2}N_{(+)}(z;\alpha)}.
\end{equation}
Similar pole structures like $\tilde{v}_{(0)}(z;\alpha)$, consisting of Markovian and non-Markovian types, can be found and have been plotted in Figure \ref{offdiagonalpoles}. There is a Markovian pole $z_{(+)}$ determined by $z_{(+)}+2g^{2}N_{(+)}\left(z_{(+)};\alpha\right)=0$, whose solution around the origin in leading order is $z_{(+)}\sim -2g^{2}N_{+}(0;\alpha)$. Two infinite sequences of non-Markovian poles are located at {$z_{(+),n}^0\sim -nH$ and $z_{(+),n}^{+}\sim-nH+i\omega$, which are in the neighborhood of the poles of $N_{(+)}(z;\alpha)$ itself. Their precise value are also determined by the equation $z+2g^2 N_{+}(z;\alpha)=0$.}

The corresponding residues at these poles are calculated as 
\begin{equation}
\left\{
\begin{aligned}
    \mathfrak{Res}\left[\tilde{v}_{(+)}(z;\alpha),z_{(+)}\right]&=\left[1-2g^{2}N_{(+)}'(0;\alpha)\right]v_{(+)}(0;\alpha)+\mathcal{O}(g^4),
    \\
    \mathfrak{Res}\left[\tilde{v}_{(+)}(z;\alpha),z_{(+),n}^0\right]&=-\frac{2g^{2}\;\mathfrak{N}_{n}\;v_{(+)}(0;\alpha)}{n^{2}H^2}+\mathcal{O}(g^4),
    \\
     \mathfrak{Res}\left[\tilde{v}_{(+)}(z;\alpha),z_{(+),n}^+\right]&=-\frac{2g^{2}\;\mathfrak{N}_{n}\;v_{(+)}(0;\alpha)}{(nH-i\omega)^{2}}+\mathcal{O}(g^4).
    \label{offdiagonalresidue}
\end{aligned}
\right.
\end{equation}
Summing up all contributions from \eqref{offdiagonalresidue} and using \eqref{InverseLaplacetransformation2}, we obtain the off-diagonal part of the detector state as (neglecting $\mathcal{O}(g^4)$ contributions)
\bea
    v_{(\pm)}(\tau;\alpha)&\equiv&v_{(\pm)}^{\text{Markovian}}+v_{(\pm)}^{\text{non-Markovian}}\no\\
    &=&v_{(\pm)}(0;\alpha)e^{-\Gamma_{(\pm)}\tau}-g^2 \bigg[S_{(\pm)}(0;\alpha)e^{-\Gamma_{(\pm)}\tau}-S_{(\pm)}(\tau;\alpha) \bigg]v_{(\pm)}(0;\alpha).
        \label{vpmt2}
\eea
where $\Gamma_{(\pm)}\equiv2g^{2}N_{(\pm)}(0;\alpha)$ and non-Markovian contribution functions $S_{(\pm)}(\tau;\alpha)$ are defined as
\begin{equation}
    S_{\pm}(\tau;\alpha)\equiv-\sum_{n=1}^{\infty}\mathfrak{N}_{n}\left(\frac{1}{n^{2}H^2}+\frac{e^{\pm i\omega\tau}}{(nH \mp i\omega)^{2}}\right)e^{-nH\tau}.
\end{equation}
Similar to the structure of \eqref{v0t2}, the functions $S_{\pm}(\tau;\alpha)$ exclusively determine all non-Markovian contributions in the non-diagonal part of the detector state, while the first term of \eqref{vpmt2} is part of the solution for the Markov-approximated master equation.

The functions $S_{(\pm)}(\tau;\alpha)$ and $\Gamma_{(\pm)}$ can be explicitly evaluated as
{
\begin{equation}
    \begin{aligned}
        S_{\pm}(\tau;\alpha)&=-\frac{1}{\pi^2\sinh{\alpha}}\left\{\frac{\cosh{\alpha}}{2}\left[e^{\pm i\omega \tau}\left(\phi\left(e^{-H\tau+ i\epsilon},\mp\frac{i\omega}{H}\right)+\phi\left(e^{-H\tau- i\epsilon},\mp\frac{i\omega}{H}\right)\right)\right.\right. \\
 -&\left.\log \left(2e^{-H\tau} (\cosh{H\tau}-\cos{\epsilon })\right)\bigg]+\left[e^{\pm i\omega t}\phi\left(-e^{-H\tau},\mp\frac{i\omega}{H}\right)-\log{\left(1+e^{-H\tau}\right)}\right]\right\},\\
  \Gamma_{(\pm)}&=\frac{\mp i\omega g^{2}}{2\pi^{2}\sinh{\alpha}}\left\{2\cosh{\alpha}\left[\log{(e^{\gamma}\epsilon)}+\psi\left(\mp\frac{i\omega}{H}\right)\right]+\psi\left(\mp \frac{i\omega}{2H}\right)-\psi\left(\frac{H\mp i\omega}{2H}\right)\right\},
  \label{coefficient2}
    \end{aligned}
\end{equation}
}
where $\psi(z)$ is the Digamma function, $\gamma$ is the Euler gamma constant, and all contributions of order $\mathcal{O}(\epsilon\log{\epsilon})$ have been neglected.

We should emphasize that the genuine off-diagonal-decay-rate for detector evolution is the real part of function $\Gamma_{(\pm)}$, which admits a decomposition as
\begin{equation}
    \Gamma_{(\pm)}=\left(\frac{\Gamma_{(0)}}{2}-\frac{g^{2}H}{2\pi^2}\coth{\frac{\alpha}{2}}\right)\mp \frac{i\Delta\omega}{2},
    \label{decomposition1}
\end{equation}
{The so-called Lamb-shifted frequency $\Delta\omega$ can be read from \eqref{coefficient2}:}
\begin{equation}
    \Delta\omega=\frac{g^{2}\omega}{\pi^2\sinh{\alpha}}\left\{2\cosh{\alpha}\left[\log{(e^{\gamma}\epsilon)}+\mathfrak{Re}\;\psi\left(\frac{i\omega}{H}\right)\right]+\mathfrak{Re}\left[\psi\left(\frac{i\omega}{2H}\right)-\psi\left(\frac{{H}+ i\omega}{2H}\right)\right]\right\},
\end{equation}
which is logarithmic divergent as $\epsilon\rightarrow 0^{+}$.

Moreover, it is interesting to note that $\Gamma_{(0)}$ is of the same order with genuine off-diagonal-decay-rate (the real part in \eqref{decomposition1}), therefore, we may use $\Gamma_{(0)}$ to encode the collective Markovian-decay-rate for the de Sitter evolution of a comoving UDW detector.

\subsection{Markovian v.s. non-Markovian dynamics}
\label{sec:Dynamics_Detector}

We {summarize} the obtained coefficient functions in Bloch decomposition \eqref{normalrho}, given by \eqref{v0t2} and \eqref{vpmt2}, as \footnote{As we mainly concern the difference between Markovian and non-Markovian decay in this subsection, we temporarily omit the de Sitter vacuum-choices parameter $\alpha$ in the expressions of functions $v_{(\mu)}$ and $S_{(\mu)}$ (with $\mu=0,\pm$) for notation simplicity.} 
\be
\left\{\begin{aligned}
v_{(0)}(\tau)   &=\gamma_{(0)}\left(e^{-\Gamma_{(0)}\tau}-1\right)+v_{(0)}(0)e^{-\Gamma_{(0)}\tau}-g^{2}\bigg[S_{(0)}(0)e^{-\Gamma_{(0)}\tau}-S_{(0)}(\tau)\bigg]v_{(0)}(0), \\
v_{(\pm)}(\tau)&=v_{(\pm)}(0)e^{-\Gamma_{(\pm)}\tau}-g^2 \bigg[S_{(\pm)}(0)e^{-\Gamma_{(\pm)}\tau}-S_{(\pm)}(\tau) \bigg]v_{(\pm)}(0).
\end{aligned}
\label{collect1}
\right.
\ee
Before employing quantum Fisher information as a quantitative probe, at the present stage, we investigate some qualitative differences between the Markovian and non-Markovian dynamics in the time-evolution \eqref{collect1} of the comoving detector.

Whenever the typical time scale of the environment is much smaller than that of the detector, we can assume the detector undergoes Markovian evolution. According to the formulation of \eqref{collect1}, this means that we can neglect all terms that include non-Markovian contribution functions $S_{(\mu)}$, leading to
\be
\left\{\begin{aligned}
v_{(0)}(\tau)   &=\gamma_{(0)}\left(e^{-\Gamma_{(0)}\tau}-1\right)+v_{(0)}(0)e^{-\Gamma_{(0)}\tau}, \\
v_{(\pm)}(\tau)&=v_{(\pm)}(0)e^{-\Gamma_{(\pm)}\tau}.
\end{aligned}
\label{collect2}
\right.
\ee
The solution satisfies a local GKSL master equation \cite{Open1}, which is a completely positive and trace-preserving (CPTP) map recasted from \eqref{REDFIELD} via Markov and RWA approximations. It is important to note that after a sufficiently long time, the decay rates $\Gamma_{(\mu)}$ drive the Bloch vector to asymptotic values. The UDW detector thus reaches a unique asymptotic state.
\be
    \rho(\tau\rightarrow{\infty})=\frac{1}{2}\begin{bmatrix}
    1-\gamma_{(0)} && 0
    \\
    0 && 1+\gamma_{(0)}
    \end{bmatrix} 
       \equiv \frac{e^{-H_{\text {detector }} / T_{\mathrm{eff}}}}{\operatorname{Tr}\left[e^{-H_{\text {detector }} / T_\mathrm{eff}}\right]}.
    \label{asymptstate}
\ee
which has a Gibbs form with an effective temperature $T_{\mathrm{eff}}$, and is irrelevant to detector initial states. In the context of a black hole background or for an accelerating detector, \eqref{asymptstate} indicates an inevitable thermalization \cite{UDW8} of the UDW detector ending at the equilibrium with Hawking-Unruh temperature \cite{Open1-1,Open1-2}. However, the de Sitter story is more involved. By substituting the coefficients \eqref{coefficient1} into \eqref{asymptstate}, we obtain
\be
    \rho(\tau\rightarrow{\infty})=
    \frac{1}{2}\begin{bmatrix}
   \frac{1+\cosh\left(\alpha+\frac{\pi\omega}{H}\right)}{1+\cosh{\alpha}\cosh{\left(\frac{\pi\omega}{H}\right)}} && 0
    \\
    0 &&  \frac{1+\cosh\left(\alpha-\frac{\pi\omega}{H}\right)}{1+\cosh{\alpha}\cosh{\left(\frac{\pi\omega}{H}\right)}}
    \end{bmatrix}+\mathcal{O}(\epsilon\log{\epsilon}),
    %    \equiv \frac{e^{-H_{\text {detector }} / T_{\mathrm{eff}}}}{\operatorname{Tr}\left[e^{-H_{\text {detector }} / T_\mathrm{eff}}\right]}
    \label{asymptstate-ds}
\ee
which means that a comoving UDW detector must approach a thermal state at Gibbons-Hawking temperature $T_{\mathrm{eff}}\equiv T_{GH}=H/2\pi$, once it is asymptotically equilibrium with background fields respecting Bunch-Davies vacuum \cite{Open1-3}. However, for general $\alpha-$vacua, even in the equilibrium state \eqref{asymptstate-ds}, we cannot conclude that the detector state is thermal in nature since it fails to satisfy the detailed balance condition \cite{dS4}. Nevertheless, in {a} unified perspective from quantum thermodynamics, we can say that the equilibrium state \eqref{asymptstate-ds} is a result of the so-called {Zeroth Law of quantum thermodynamics} characterized by a relative entropic formulation \cite{Open7}.

From a general open dynamics \eqref{collect1}, {we can observe that the non-Markovian contributions become more significant at early time}, which means $H\tau \lesssim 1$ for de Sitter context. However, during late-time evolution with $H\tau\gg 1$, the non-Markovian contributions can be absorbed into the initial state preparation of the detector, i.e., every general solution \eqref{collect1} looks like a Markov-type solution for sufficiently large $\tau$, except it has {altered} initial state different from the one for explicit Markovian solution \eqref{collect2}.

This claim can be checked by investigating the explicit solutions \eqref{v0t2} and \eqref{vpmt2}. {The real parts of the non-Markovian poles satisfy}
\begin{equation}
   \mathfrak{Re}\left[z\right]\sim-nH+\mathcal{O}(g^{2}),\quad n>0.
\end{equation}
At late time $H\tau\gg 1$, the non-Markovian contribution {scales as} $S_{(\mu)}(\tau)\sim \mathcal{O}(e^{-H\tau})$ for $\mu=0,\pm$, which gives
\begin{equation}
\left\{
    \begin{aligned}
        v_{(0)}(\tau;\alpha)&=\gamma_{(0)}\left(e^{-\Gamma_{(0)}\tau}-1\right)+\left(\left[1-g^{2}S_{(0)}(0;\alpha)\right]\left.v_{(0)}\right|_{\tau=0}\right)e^{-\Gamma_{(0)}\tau}+g^{2}\mathcal{O}(e^{-H\tau}),
        \\
        v_{(\pm)}(\tau;\alpha)&=\left(\left[1-g^{2}S_{(0)}(0;\alpha)\right]\left.v_{(\pm)}\right|_{\tau=0}\right)e^{-\Gamma_{(\pm)}\tau}+g^{2}\mathcal{O}(e^{-H\tau}).
        \label{shiftedmarkov}
    \end{aligned}
    \right.
\end{equation}
If we {regard}
\begin{equation}
    \begin{aligned}
        \left.v'_{(0)}\right|_{\tau=0}&\equiv\left[1-g^{2}S_{(0)}(0;\alpha)\right]\left.v_{(0)}\right|_{\tau=0},\\
        \left.v'_{(\pm)}\right|_{\tau=0}&\equiv\left[1-g^{2}S_{(0)}(0;\alpha)\right]\left.v_{(\pm)}\right|_{\tau=0}.
    \end{aligned}
\end{equation}
{as shifted initial conditions}, \eqref{shiftedmarkov} {are in} the same form as {the Markovian solution} \eqref{collect2}. Therefore, we refer the detector density matrix corresponding to \eqref{shiftedmarkov} as a {shifted-Markovian solution} with altered initial conditions $v_{(0)}'(0)$ and $v_{(\pm)}'(0)$. Since $\mathfrak{Re}\left[S_{(\mu)}(0)\right]>0$ by positivity, one notes that $|v_{(\mu)}'(0)|<|v_{(\mu)}(0)|$, which is useful {for numerical analysis}. 

\begin{figure}[htbp]
\centering
\includegraphics[width=.7\textwidth]{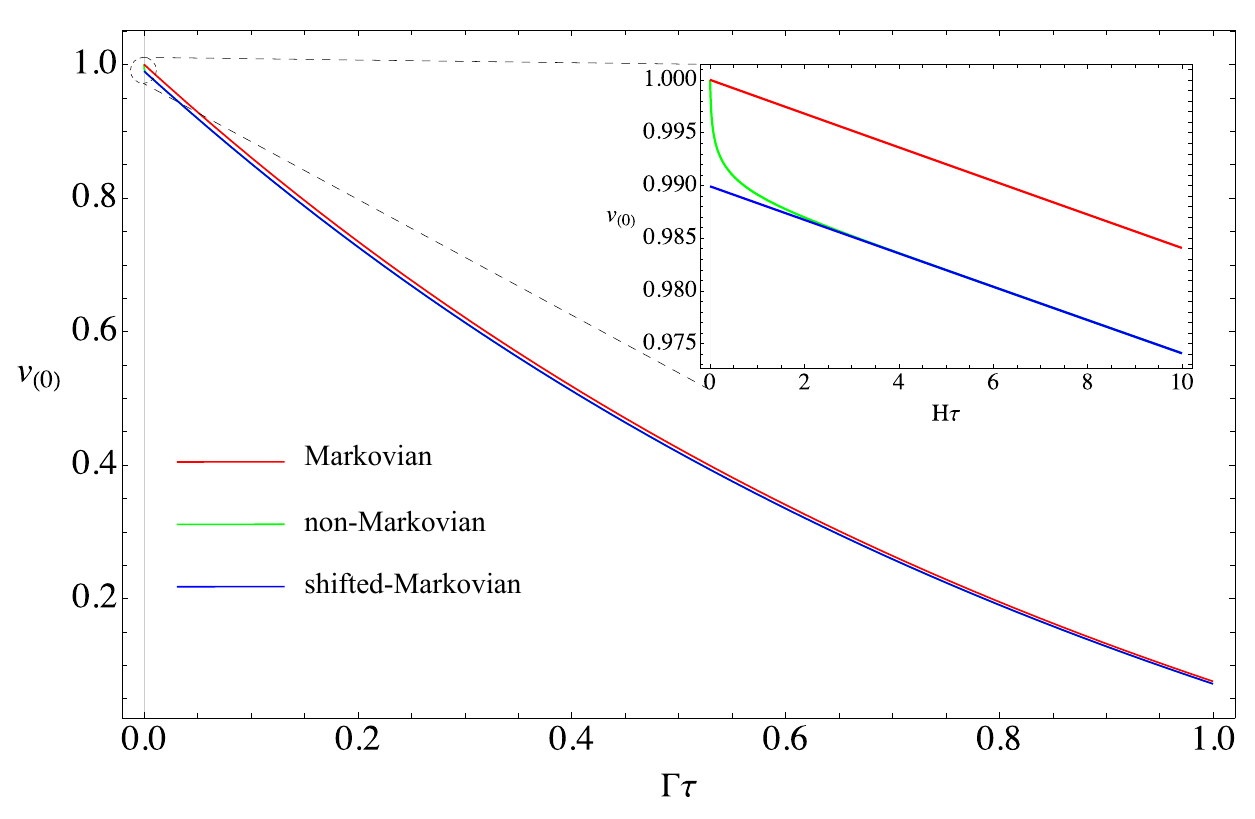}
\caption{The early- and late-time behaviors of $v_{(0)}(\tau)$ for Markovian, non-Markovian and Shifted Markovian solutions. {The detector evolution starts from} $\rho(0)=|1\rangle\langle 1|$, with $g=0.1$, $\omega=1$, $H=2\pi$, {\red$\epsilon=10^{-3}$}, {and the background are chosen as in Bunch-Davies vacuum, i.e., $\alpha\rightarrow-\infty$.} \label{NMMDAM100}}
\end{figure}

In Figure~\ref{NMMDAM100}, we illustrate the evolution of the diagonal part of the detector density matrix, say $v_{(0)}=\rho_{11}-\rho_{00}$, undergoing either Markovian or non-Markovian evolutions. One can observe that the difference between the two types of dynamics mainly appears at the early time of the detector evolution (depicted in the inset of Fig.~\ref{NMMDAM100}). As time grows, the non-Markovian contributions decay rapidly for $H\tau\lesssim 1$ and approach the Shifted Markovian solution for $H\tau\gg 1$, which decays exponentially for large $\tau$.

\begin{figure}[htbp]
\centering
\subfloat[$H=2\pi\ \text{(solid)},5\pi\ \text{(dashed)},15\pi\ \text{(dash-dotted)}$]{\includegraphics[width=.5\textwidth]{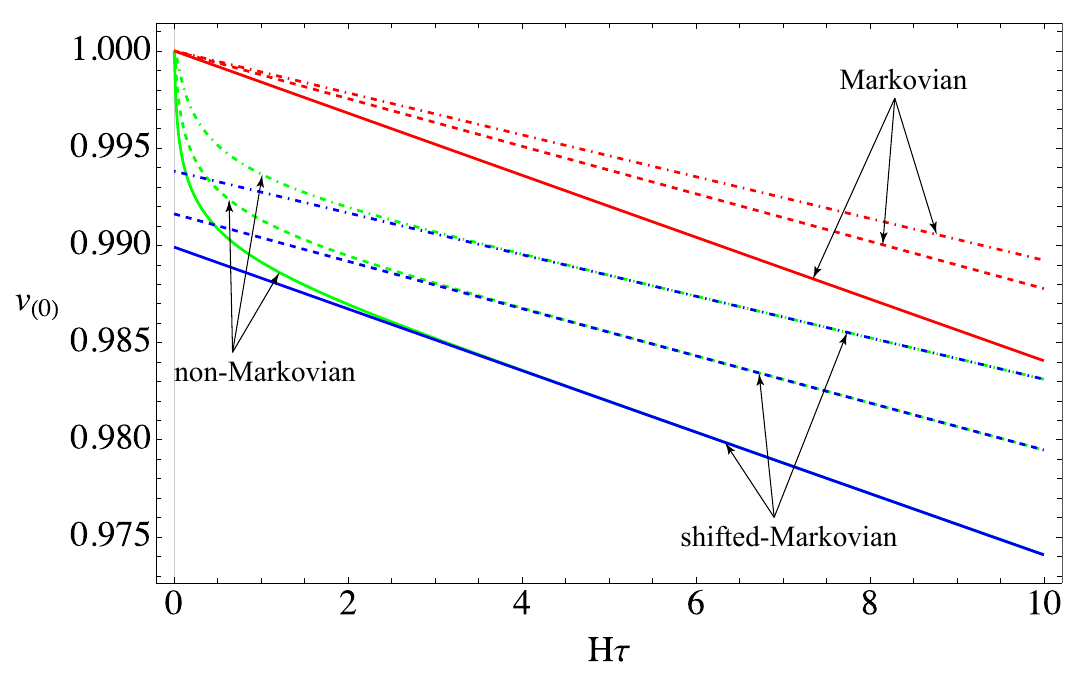}}~
\subfloat[$\alpha=-\infty\ \text{(solid)},-2\ \text{(dashed)},-1\ \text{(dash-dotted)}$]{\includegraphics[width=.5\textwidth]{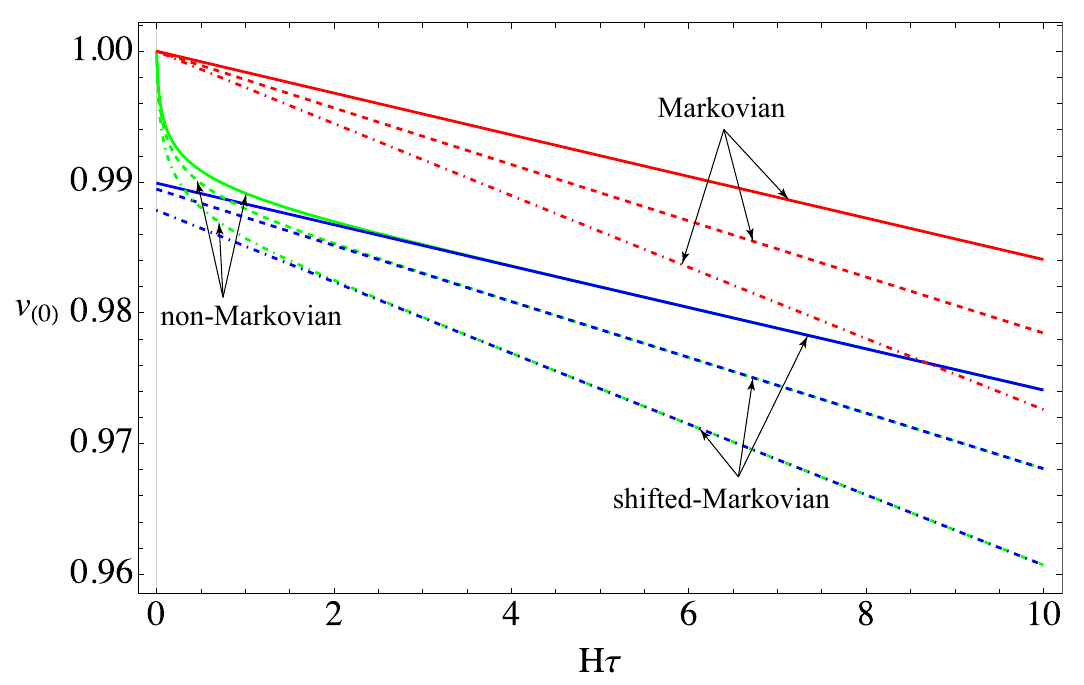}}
\caption{The early-time behavior of $v_{(0)}(\tau)$ for Markovian, non-Markovian, and shifted-Markovian solutions is analyzed. The detector starts from the initial state $\rho(0)=|1\rangle\langle 1|$, with $g=0.1$, $\omega=1$ {and $\epsilon=10^{-3}$}. In (a), the Hubble parameter is valued as $H=2\pi$ (solid lines), $5\pi$ (dashed lines), and $15\pi$ (dash-dotted lines), with the Bunch-Davies vacuum chosen, i.e., $\alpha\rightarrow-\infty$. With larger Hubble parameter or Gibbons-Hawking temperature increases, the differences between the three solutions decrease. In (b), the calculations are performed with a fixed Hubble parameter $H=2\pi$ but for different de Sitter vacua, such as $\alpha=-\infty$ (solid lines), $-2$ (dashed lines), and $-1$ (dash-dotted lines). With larger deviation from the Bunch-Davies vacuum, the differences among the three solutions increase. }
\label{fig4}
\end{figure}

Finally, we would like to further investigate how the spacetime geometry (Hubble parameter $H$) and vacuum-choices (superselection parameter $\alpha$) would invoke the difference between Markovian, non-Markovian, and Shifted Markovian solutions for the UDW detector. In Figure~\ref{fig4}(a), the early-time behavior of $v_{(0)}$ within the chosen Bunch-Davies vacuum has been estimated for specific values of the Hubble parameter ($H=2\pi,5\pi,15\pi$). We observe that as the Hubble parameter increases, the difference between Markovian, non-Markovian, and shifted-Markovian solutions decreases. Keeping in mind that the comoving detector in de Sitter should be immersed in Gibbons-Hawking radiation at a temperature $T_{GH}=H/2\pi$, we suggest that the initial difference between the three dynamics has been smoothed out by a hotter environment bath. 

In Figure~\ref{fig4}(b), a similar calculation of $v_{(0)}$ has been conducted with a fixed Hubble parameter $H=2\pi$ but for general de Sitter $\alpha-$vacua. We note that as the deviation from the Bunch-Davies choice ($\alpha=-\infty$) increases, the initial difference between Markovian and shifted-Markovian solutions escalates significantly. In other words, within non-Bunch-Davies vacua in de Sitter, the non-Markovianity of detector evolution is enhanced. Nevertheless, this may not be so surprising if we admire the suggested role played by general $\alpha-$vacua in the context of quantum gravity \cite{dS8+}. That is, the superselection parameter $\alpha$ can somehow be related to a minimal length $e^{\alpha}\sim H/\Lambda$ where fundamental new physics may emerge. Certainly, the Markovian approximation should no longer be reliable in such an extreme case \cite{NonM1}.

\subsection{Renormalization of the Evolution}
We have derived the solution \eqref{collect1} for the detector dynamics. However, there is an explicit dependence on $\epsilon$ exhibited, which is undesirable as $\epsilon$ is tied to the cutoff of the background fields. In this section, we explore the procedure for eliminating the $\epsilon$-dependence and reformulating the final results solely in terms of physical parameters.
\subsubsection{Markovian Case}
As a starting point, we first discuss the renormalization of Markovian evolution \eqref{collect2}. The only divergence in the expression comes from the logarithmically divergent Lamb shift $\Delta \omega$. By choosing the decomposition $\omega=\Omega-\delta\omega$ (where $\Omega$ is the shifted physical frequency and $\delta\omega$ combines the Lamb shift and any counterterm for simplicity), we may express the solution \eqref{collect2} in terms of the renormalized physical quantities only. 

Since $\omega$ is the bare coupling, it is no longer suitable for discussion in the interaction picture. Switching back to Schr\"odinger's picture, we have
\be
\left\{\begin{aligned}
v_{(0)}(\tau)   &=\gamma_{(0)}^P\left(e^{-\Gamma_{(0)}^P\tau}-1\right)+v_{(0)}(0)e^{-\Gamma_{(0)}^P\tau}, \\
v_{(\pm)}(\tau)&=v_{(\pm)}(0)e^{-\Gamma_{(\pm)}^{P}\tau},
\end{aligned}
\right.
\ee
where the superscript $P$ denotes that the quantities have been "renormalized" by replacing $\omega$ with the physical frequency $\Omega$: 
\begin{equation}
\left\{
    \begin{aligned}
        \gamma_{(0)}^{P}&=-\frac{\sinh{\alpha}\sinh{\left(\frac{\pi\Omega}{H}\right)}}{1+\cosh{\alpha}\cosh{\left(\frac{\pi\Omega}{H}\right)}},\\
        \Gamma_{(0)}^{P}&=-\frac{g^{2}\Omega}{\pi}\left[\frac{\sinh{\alpha}\sinh{\left(\frac{\pi\Omega}{H}\right)}}{1+\cosh{\alpha}\cosh{\left(\frac{\pi\Omega}{H}\right)}}\right]^{-1},\\
        \Gamma_{(\pm)}^{P}&=\left(\frac{\Gamma_{(0)}^P}{2}-\frac{g^{2}H}{2\pi^2}\coth{\frac{\alpha}{2}}\right)\mp \frac{i\Omega}{2}.
    \end{aligned}
    \right.
\end{equation}
Since $\delta\omega\sim \mathcal{O}(g^2)$, we note that replacing with $\omega$ by $\Omega$ for $\gamma_{(0)}^p$ and $\Gamma_{(0)}^p$ only leads to $\mathcal{O}(g^4)$ contributions in the final results. 

\subsubsection{non-Markovian Case}
Renormalization of the non-Markovian solution is inherently more challenging. For \eqref{collect1}, in addition to the divergent Lamb shift $\Delta\omega$, we must address that the logarithmic divergences arising also from $S_{(0)}(0)$ and $S_{(\pm)}(0)$. As mentioned in discussion of \eqref{shiftedmarkov}, $S_{(0)}(0)$ and $S_{(\pm)}(0)$ are linked to shifts in the initial conditions $v_{(0)}(0)$ and $v_{(\pm)}(0)$ due to non-Markovian effects. Consequently, expressing the final results in terms of $v_{(0)}(0)$ and $v_{(\pm)}(0)$ inevitably retains these divergences. 

Nevertheless, this does not preclude renormalization; instead, it reflects the sensitivity of the detector dynamics at early times ($\tau \sim \epsilon$) to the cutoff. This is quite natural in the sense that the energy scale probed by the detector should scale as $E \sim 1/\tau$ for small $\tau$.

To renormalize the non-Markovian dynamics, we could further fix $v_{(\mu)}(\tau)$ at a small but finite physical time scale $\tau_0 = \kappa$, and then express $v_{(\mu)}(\tau)$ in terms of $v_{(\mu)}(\kappa)$. This procedure reflects the selection of a renormalization condition at a fixed external energy scale in field theory. For convenience, we decompose the time variable as $\tau =\tau_0+t= \kappa + t$. Since $\kappa$ is assumed to be small, we focus primarily on the regime $t > 0$. From \eqref{collect1}, we obtain non-Markovian solution in the Schr\"{o}dinger picture (for details see Appendix \ref{appendixE}):

\be
\left\{\begin{aligned}
v_{(0)}(t)   &=\gamma_{(0)}^P\left(e^{-\Gamma_{(0)}^Pt}-1\right)+v_{(0)}(\kappa)\Big[1+g^2\left(S_{(0)}(t+\kappa,\Omega)-S_{(0)}(\kappa,\Omega)\right)\Big]e^{-\Gamma_{(0)}^Pt}, \\
v_{(\pm)}(t)&=v_{(\pm)}(\kappa)\Big[1+g^2\left(S_{(\pm)}(t+\kappa,\Omega)-S_{(\pm)}(\kappa,\Omega)\right)\Big]e^{-\Gamma_{(\pm)}^P t},
\label{Re_non_Markov}
\end{aligned}
\right.
\ee
Note that we have used the abbreviation $v(t)\equiv v(t+\kappa)$ and $S_{(\mu)}(\tau,\Omega)$ instead of $S_{(\mu)}(\tau)$ to highlight the fact that we should replace $\omega$ with the physical frequency $\Omega$ in the function $S_{(\mu)}(\tau)$. This discussion completes the renormalization issue of the non-Markovian evolution of a qubit.

\section{Quantum Fisher information of a comoving UDW detector}
\label{sec:QFI}

As mentioned earlier, to fully describe the thermalization process of the comoving UDW detector, we need to employ some efficient process functions. We choose quantum Fisher information, a crucial concept in quantum metrology and information geometry \cite{QFI1-4,QFI1-5}. The aim of this section is to explore the QFI of the detector's non-Markovian evolution given by \eqref{normalrho} and \eqref{collect1}, while the Hubble parameter $H$ serves as the parameter to be optimally estimated. We are concerned with how the non-Markovian contribution and the influence from general de Sitter vacuum-choices can be recognized from QFI, especially regarding its early-time behavior as implied by the results of Section \ref{sec:Dynamics_Detector}. Within a metrological task on estimating spacetime geometry \cite{Open8-5}, QFI bounds the maximal precision of any estimator. Therefore, it is also interesting to investigate how to effectively enhance the QFI by designing a UDW detector, e.g., careful selection of its initial state, level spacing, etc. Finally, as the equilibrium state \eqref{asymptstate} depends solely on Gibbons-Hawking temperature and vacuum-choices $\alpha$, we would naturally expect that the asymptotic behavior of QFI should manifest the thermal nature of de Sitter space.

\subsection{Definitions and properties}
\label{sec:Def_QFI}
Quantum Fisher information (QFI) is related to the parameter estimating problem in quantum metrology  \cite{QFI1-1,QFI1-2,QFI1-3}. Classically, for a $F$-valued random variable $R_{X}:\Omega \rightarrow F$ that depends on a real parameter $X\in \mathbb{R}$, the parameter estimating problem discusses the estimation of $X$ through measuring $R_{X}$.  An estimator for sample size $N$ is a map $\hat{X}:F^{N}\rightarrow \mathbb{R}$, which estimates $X$ based on the results of measuring $R_{X}$ for $N$ times. The estimator is unbiased if  $E[\hat{X}]=X$. For the random variable $R_{X}$, the classical Fisher information is defined as 
\begin{equation}
    \mathcal{F}_{C}(X)\equiv\int\mathrm{d}s\,p(s;X)\left(\frac{\partial \log{p(s;X)}}{\partial X}\right)^2.
\end{equation}
For an unbiased estimator $\hat{X}$ with sample size $N$, the Cram\'{e}r-Rao bound
\begin{equation}
    \mathrm{Var}(\hat{X})\geq\frac{1}{N\mathcal{F}_C(X)},
    \label{Cramer}
\end{equation}
relates  $\mathcal{F}_C(X)$ to the variance of the estimator $\hat{X}$. As a result, the classical Fisher information $\mathcal{F}_C(X)$ essentially represents the amount of information about $X$ that one may extract through measuring $R_{X}$.

In the quantum regime, instead of random variables $R_{X}$, one deals with quantum states $\rho_{X}$. Measuring any observable $\hat{A}$ over the Hilbert space $\mathcal{H}$ generates a parameter-dependent random variable $A_{X}$, for which the variance is bounded by \eqref{Cramer}. Defining operator $L$ over the Hilbert space for which the condition
\begin{equation}
    \partial_{X}\rho=\frac{1}{2}\{L,\rho\},
\end{equation}
is satisfied. The QFI is defined as
\begin{equation}
    \mathcal{F}_{Q}(X)\equiv\mathrm{Tr}[\rho L^{2}].
\end{equation}
In terms of eigenvectors and eigenvalues of the density matrix $\rho(X)$, we have \cite{QFI1-2,QFI1-3}
\begin{equation}
    \mathcal{F}_{Q}(X)=\sum_{i} \frac{(\partial_{X}\lambda_{i})^{2}}{\lambda_{i}}+\sum_{i,j}\frac{2(\lambda_{i}-\lambda_{j})^{2}}{\lambda_{i}+\lambda_{j}}|\langle \psi_{i},\partial_{X}\psi_{j}\rangle|^{2}.
    \label{FisherQ}
\end{equation}
For any observable $\hat{A}$, we have $\mathcal{F}_{C}(X,A)\leq \mathcal{F}_{Q}(X)$. Therefore, for the quantum estimation problem, we have
\begin{equation}
    \mathrm{Var}(\hat{X})\geq\frac{1}{N\mathcal{F}_Q(X)}.
    \label{CramerQM}
\end{equation}
That is, similar to the role of $\mathcal{F}_C(X)$ in classical measurements, $\mathcal{F}_Q(X)$ qualifies the amounts of information that we can extract about $X$ by measuring quantum states $\rho_{X}$. 

For a general Hilbert space, \eqref{FisherQ} is rather complicated for calculation. However, for $\mathrm{dim}\,\mathcal{H}=2$, the qubit density matrix admits a Bloch decomposition; therefore the related QFI can be given explicitly  (e.g., from \eqref{rho}) as
\begin{equation}
    \mathcal{F}_{Q}(X)=\sum_{\mu=0,\pm}\left\{\left[\partial_{X} {v}_{(\mu)}\right]^2+\frac{\left({v}_{(\mu)}\partial_{X} {v}_{(\mu)}\right)^{2}}{1-|\vec{v}|^{2}}\right\},
    \label{QFI-1}
\end{equation}
which is more convenient for numerical purposes.

For our purpose, we aim to estimate the Hubble parameter $H$, which characterizes de Sitter geometry. For a comoving UDW detector undergoing general non-Markovian dynamics, the related QFI $\mathcal{F}_{Q}(H;\alpha)$ can be numerically estimated by substituting \eqref{Re_non_Markov} into \eqref{QFI-1}, where the superselection parameter $\alpha$ labels different de Sitter vacua.

\subsection{QFI for the detector in de Sitter Space}
\label{QFIindS}
\subsubsection{Early-time analysis}
\begin{figure}[htbp]
\centering
\includegraphics[width=.7\textwidth]{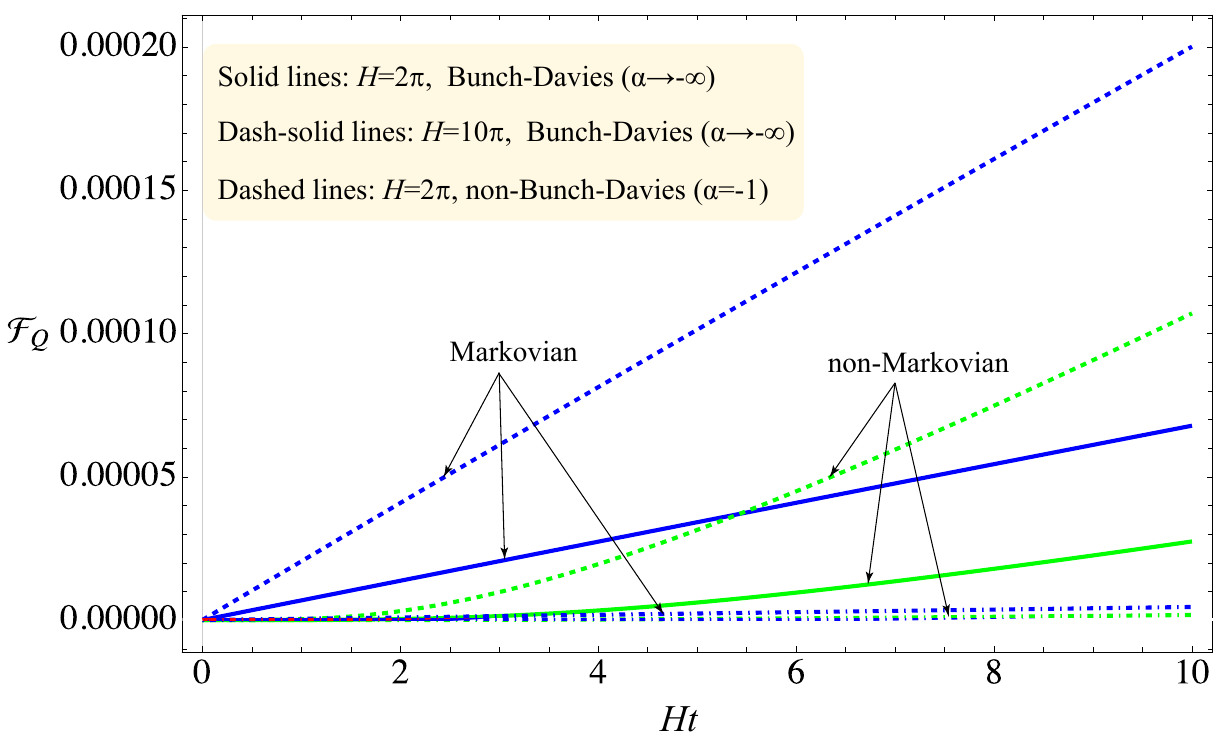}
\caption{Early-time behavior of $\mathcal{F}_{Q}(H;\alpha)$ for Markovian  and non-Markovian solutions. {The detector initial condition is $\rho(0)=|1\rangle\langle 1|$, with $g=0.1$, {$\Omega=1$ and $\kappa H=10^{-3}$}}. For the detector undergoing non-Markovian dynamics, the cooler environment bath and larger deviation from Bunch-Davies vacuum lead to a larger $\mathcal{F}_{Q}(H;\alpha)$ for Hubble parameter estimation. 
\label{NMMFAS100}}
\end{figure}

We begin by analyzing the early-time behavior of $\mathcal{F}_{Q}(H;\alpha)$. As the natural dimensionless time scale in the short time regime is {$Ht$}, we plot $\mathcal{F}_{Q}(H;\alpha)$ with respect to $Ht$ in Figure~\ref{NMMFAS100}, where the detector initial state has been set to $\rho(0)=|1\rangle\langle 1|$. We compare the early-time behaviors of QFI related to non-Markovian and Markovian solutions, respectively. 

{Firstly, we observe that  $\mathcal{F}_{Q}(H;\alpha)$ exhibits linear growth for Markovian solutions, while the growth for non-Markovian solutions is significantly slower. In particular, we find that $\mathcal{F}_{Q}(H;\alpha)\approx 0$ for non-Markovian solutions when $Ht \lesssim 1$. This is quite general owing to the early-time behavior of the correlation function $\Delta^{\pm}(\tau)$. For simplicity, we take $\alpha=-\infty$, set $\epsilon':=\epsilon /H$ as the cutoff with a length/time scale, we have
\begin{equation}
    \Delta^{+}(\tau)|_{\alpha=-\infty}=-\frac{H^{2}}{16\pi^{2}\sinh^{2}\big(H(\tau-i\epsilon')/2\big)}=-\frac{1}{4\pi^2(\tau-i\epsilon')^2}(1+O(H^2\tau^2)).
    \label{DeltaBunchDavies}
\end{equation}
which indicates that for $H\tau\ll 1$, the particle can not distinguish the background from that in the Minkowskian case. Therefore, the exact evolution of the detector at very early times is independent of $H$, and one can not tell any useful information about $H$ from $\rho(\tau)$, i.e., $\mathcal{F}_{Q}(H)$ is nearly vanishing. This agrees with the results of non-Markovian solutions.  However, when the Markov approximation is applied,  the information at early times is lost and we only observe a linear growth of $\mathcal{F}_{Q}(H)$. This indicates that the Markovian approximation is unreliable in the early-time era of the evolution.}

Secondly, we note that while the QFI associated with non-Markovian dynamics is consistently lower than that linked to the Markovian solution, the discrepancy between them depends on the Hubble parameter $H$ and vacuum choices $\alpha$. In particular, at larger Hubble parameters $H$ (or higher Gibbons-Hawking temperatures $T_{GB}$), the gap between the non-Markovian $\mathcal{F}_{Q}(H;\alpha)$ and the one related to Markovian solution diminishes, illustrating our earlier observation in Figure \ref{fig4}(a) that distinct dynamics exhibit similar behavior in a hotter environment. On the other hand, for the background field situated in non-Bunch-Davies vacua, the difference between non-Markovian and Markovian dynamics can be recognized more easily, since the deviation in the related $\mathcal{F}_{Q}(H;\alpha)$ significantly intensifies as $\alpha\rightarrow0$. This further supports the findings shown in Figure \ref{fig4}(b). 

\begin{figure}[htbp]
\centering
\includegraphics[width=.8\textwidth]{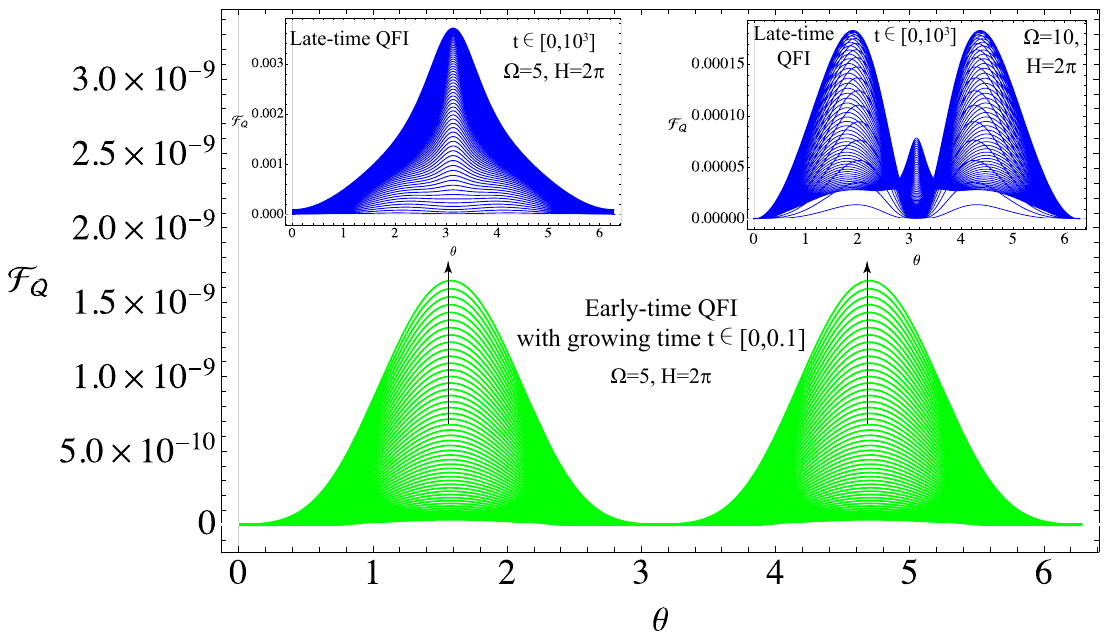}
\caption{The early and late time behaviors of $\mathcal{F}_{Q}(H;\alpha)$ for arbitrary detector initial states $\rho(0)=|\psi_{\theta}\rangle\langle\psi_{\theta}|$ with  $g=0.1$, $H=2\pi$, {$\kappa H=10^{-3}$} and $\alpha=-\infty$. Each curve (green or blue) are plotted at a specific proper time $\tau$. Along the arrow lines, the time $\tau$ increases monotonously. \label{NMFS100}}
\end{figure}

From a metrological perspective, we are interested in reaching an optimal measurement accuracy of physical parameters, which is quantified by extreme values of QFI. From Figure \ref{NMMFAS100}, for the detector undergoing non-Markovian dynamics, we observe that the cooler environment bath and larger deviation from Bunch-Davies vacuum could lead to a larger $\mathcal{F}_{Q}(H;\alpha)$ for Hubble parameter estimation. Nevertheless, beside its monotonously increasing at early time, to determine the extreme values of $\mathcal{F}_{Q}(H;\alpha)$, we still have to investigate the late-time behavior of QFI for arbitrary initial detector states (say, $|\psi_{\theta}\rangle=\cos{\frac{\theta}{2}}|1\rangle+\sin{\frac{\theta}{2}}|0\rangle$\footnote{{We do not explore the initial states with phase parameter like $\cos \frac{\theta}{2}|1\rangle+e^{i \varphi}\sin \frac{\theta}{2} |0\rangle$, because that the QFI with respect to the Hubble parameter is generally independent of the phase $\varphi$ as shown in \cite{Open8-2}.}}). 

For arbitrary detector initial states $\rho(0)=|\psi_{\theta}\rangle\langle\psi_{\theta}|$, we depict both early- and late-time behaviors of $\mathcal{F}_{Q}(H;-\infty)$ in Figure \ref{NMFS100}. The green curves indicate the early-time QFI, which is periodic with $\theta$ and exhibits maximal values for $\theta=(n+1/2)\pi$ ($n\in\mathbb{N}$). Numerically, we confirm that this is a robust and stable pattern for varying {renormalized $\Omega$} or $H$. On the other hand, things get involved at late-time regions. we note that different patterns of QFI curves may exist by tuning $\Omega$ {and} $H$. For example, the left inset of Figure \ref{NMFS100} shows a single maximum of QFI curves located for the detector initial state prepared with $\theta=n\pi$, while the right inset with larger {renormalized $\Omega$} shows two maxima of QFI exist. 
 
To conclude this subsection, we emphasize another important observation from Figure \ref{NMFS100}. When $t$ is sufficiently large, it can be noted that the QFI curves (blue curves in the two insets) become intensive and eventually condense to a well-defined boundary. This indicates that, after a sufficiently long time, $\mathcal{F}_{Q}(H;\alpha)$ for arbitrary {renormalized $\Omega$} and $H$ would asymptotically approach a fixed value. This indeed supports the notion that the comoving UDW detector in de Sitter should reach a unique equilibrium state, corresponding to a fixed QFI.

\subsubsection{Late-time analysis}
In this section, we further explore the late-time behavior of $\mathcal{F}_{Q}(H;\alpha)$. Since it exhibits a unique asymptotic value for sufficiently late times, we are interested in how the detector approaches this thermalization end, as demonstrated by the time evolution of $\mathcal{F}_{Q}(H;\alpha)$. In other words, the core issue of this section is to use $\mathcal{F}_{Q}(H;\alpha)$ to distinguish various thermalization paths in the detector Hilbert space leading to a unique end.

We start by working with a detector in the Bunch-Davies vacuum. For a fixed Hubble parameter, the related QFI $\mathcal{F}_{Q}(H;-\infty)$ is determined by the delicate design of the detector, i.e., the energy spacing $\omega$ and initial state preparation $\theta$. In Figure~\ref{NMMFL100}, we show the behaviors of $\mathcal{F}_{Q}(H;-\infty)$ for various choices of $\Omega$ and $\theta$, while the time evolution under non-Markovian (Markovian) dynamics is depicted by solid (dashed) lines. The sub-figures in Figure~\ref{NMMFL100} correspond to specifically designed energy spacings $\Omega=0.25,1,3,6,10,15$, while the colored $\mathcal{F}_{Q}(H;-\infty)$ curves in each sub-figure are determined by different initial states $\theta=0,\pi/4,\pi/2,3\pi/4,\pi$. {In this sense, we can say that QFI distinguishes the different thermalization paths of the detector.

\begin{figure}[htbp]
\centering
\includegraphics[width=1.02\textwidth]{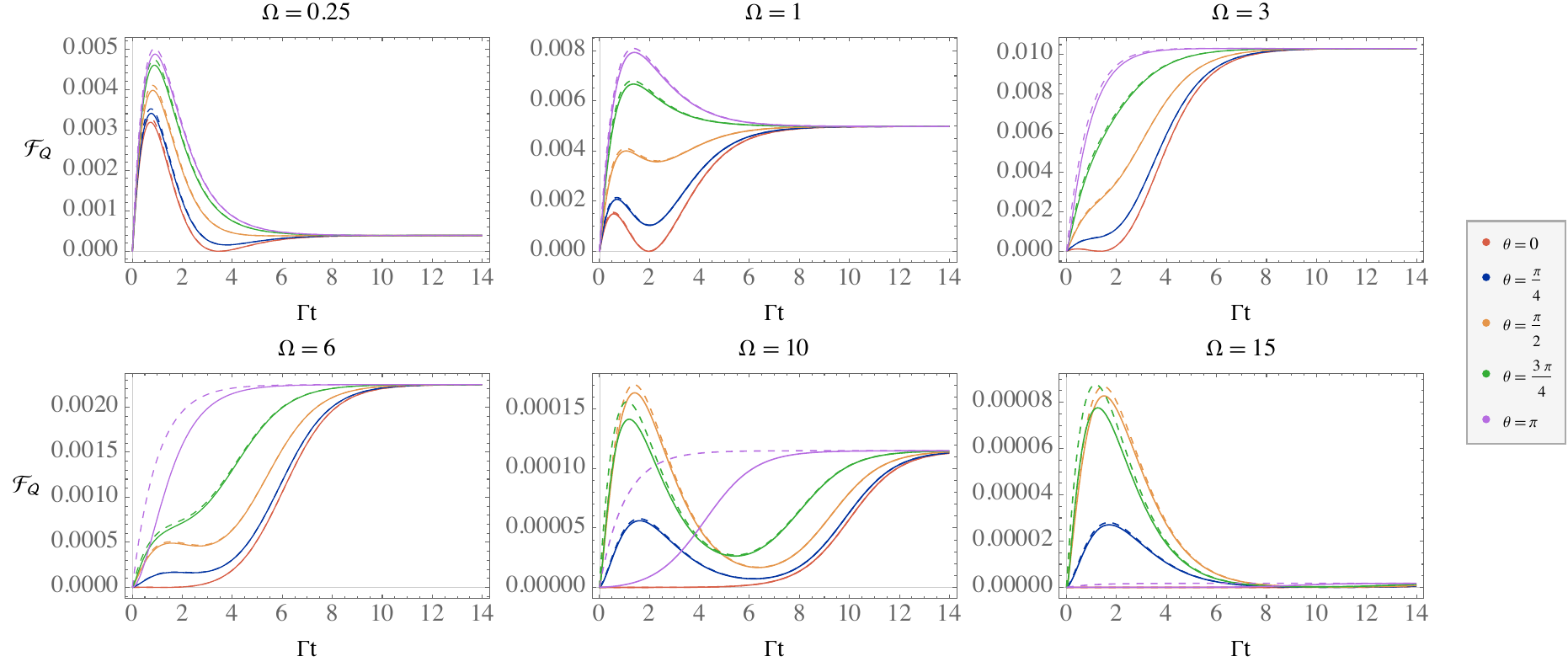}
\caption{Late-time behavior of $\mathcal{F}_{Q}(H;-\infty)$ for Markovian {(dashed lines)} and non-Markovian {(solid lines)} solutions. {The numerical analysis is performed for $g=0.1$, $H=2\pi$ {and $\kappa H=10^{-3}$}, with various detector energy spacing ($\Omega=0.25,1,3,6,10,15$) and initial states ($\theta=0,\pi/4,\pi/2,3\pi/4,\pi$)}. \label{NMMFL100}}
\end{figure}

It is obvious that after a sufficiently long time, the detector QFI under either dynamics will converge to a unique end, which is irrelevant to the initial state preparation $\theta$, and manifests the zeroth law of quantum thermodynamics for detector thermalization in de Sitter space. For a fixed Hubble parameter, the specific value of the converging end relies on the {renormalized frequency $\Omega$}. One can see that for extremely small or large $\Omega$, the QFI $\mathcal{F}_{Q}(H;-\infty)$ asymptotically converges to zero. This can be confirmed by the detector's equilibrium state \eqref{asymptstate-ds} in the long-time limit. With Bunch-Davies choice $\alpha=-\infty$, it yields
\be
    \rho^{BD}_{\text{asym}}{=}
    \frac{1}{2}\begin{bmatrix}
 1-\tanh\left(\frac{\pi\Omega}{H}\right) && 0
    \\
    0 &&  1+\tanh\left(\frac{\pi\Omega}{H}\right)
    \end{bmatrix}.
    \label{asympstateBD}%    \equiv \frac{e^{-H_{\text {detector }} / T_{\mathrm{eff}}}}{\operatorname{Tr}\left[e^{-H_{\text {detector }} / T_\mathrm{eff}}\right]}
\ee
For extremely small $\Omega$, we have a $H$-independent asymptotic state $\rho^{BD}_{\text{asym}}\approx\mathbb{I}/2$, leading to a vanishing QFI for Hubble parameter estimation. On the other hand, {for large $\Omega$}, the density matrix approaches $\rho^{BD}_{\text{asym}}\approx|0\rangle\langle0|/2$, which is again independent of $H$. In conclusion, the maximal asymptotic QFI can only be achieved for fine-tuned detector energy gap $\Omega$, {which leads to a final state \eqref{asympstateBD} that acquires most $H$-dependence.}

Moreover, we observe from Figure~\ref{NMMFL100} that during the time evolution, the QFI related to non-Markovian dynamics is consistently smaller than in the case of Markovian evolution. Recall from section \ref{sec:Dynamics_Detector} that non-Markovian dynamics resembles a shifted-Markovian solution in the late-time regime; therefore, the shift of the detector's initial condition always reduces the coherence of the density matrix.

\begin{figure}[htbp]
\centering
\includegraphics[width=1.0\textwidth]{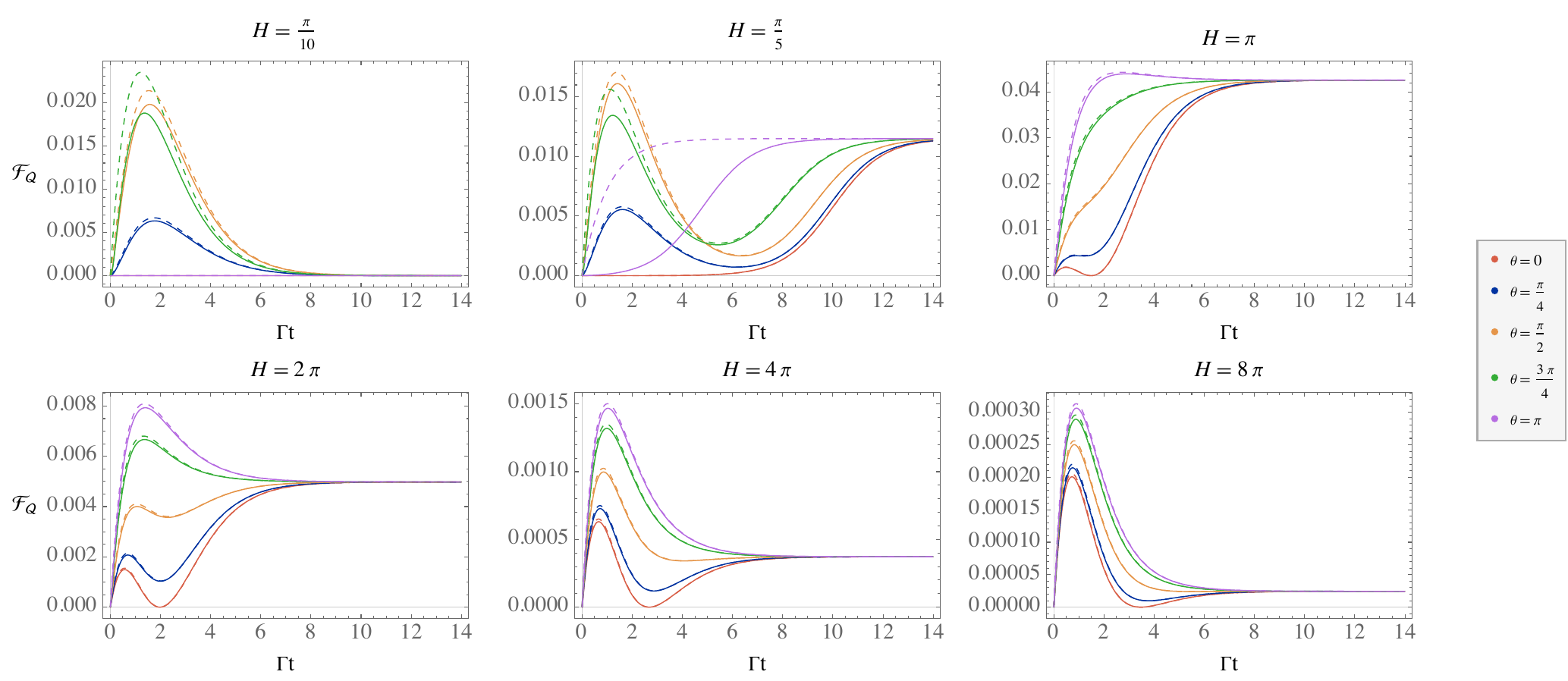}
\caption{Late-time behavior of $\mathcal{F}_{Q}(H;-\infty)$ for Markovian {(dashed lines)} and non-Markovian {(solid lines)} solutions. The estimation has been taken under $g=0.1$, {$\Omega=1$ and $\kappa H=10^{-3}$}, for various Hubble parameter values (with $H=0.1\pi,0.2\pi,\pi,2\pi,4\pi,8\pi$) and initial states (with $\theta=0,\pi/4,\pi/2,3\pi/4,\pi$). \label{fig8}}
\end{figure}

We now turn to the case of the detector with a fixed energy gap but estimating a varying Hubble parameter. Intuitively, with optimal measurement, it could be easier to measure a large parameter. However, as shown in Figure~\ref{fig8}, one can observe that for a large Hubble parameter, the related asymptotic QFI may approach zero, exhibiting a similar pattern of time evolution to that shown in Figure~\ref{NMMFL100}. This may not be surprising since the Hubble parameter and the detector energy gap are always combined as ${\Omega}/{H}$ in \eqref{asympstateBD}. Nevertheless, upon considering the two-fold role of the Hubble parameter, i.e., as a parameter to be estimated metrologically and as a determinant of a Gibbons-Hawking bath, we gain a more physical interpretation of the vanishing asymptotic QFI for large $H$, as it means a hotter de Sitter bath that disturbs any optimal quantum measurement of the Hubble parameter.

We would like to explore how the de Sitter vacuum-choices affect the time-evolution of the QFI for Hubble parameter estimation. The numerical estimation of the time evolving $\mathcal{F}_{Q}(H;\alpha)$ is shown in Figure~\ref{aNMMFL100}. Similar to the Bunch-Davies choice, for general $\alpha-$vacua, the QFI $\mathcal{F}_{Q}(H;\alpha)$ undergoing Non-Markovian dynamics remains lower than its Markovian counterpart. Moreover, we observe that for a fixed Hubble parameter and detector energy gap, the further the de Sitter vacuum deviates from the Bunch-Davies choice, the lower asymptotically converging end of $\mathcal{F}_{Q}(H;\alpha)$ can be obtained. Such suppression can be attributed to the emerging $\alpha-$dependence in the detector equilibrium state \eqref{asymptstate-ds} for the choices of non-Bunch-Davies vacuum in de Sitter field theory. As $\alpha$ approaches $0^-$, we have $\gamma_{(0)}\rightarrow 0$, leading to a $H$-independent equilibrium state $\rho^{non-BD}_{\text{asym}}\approx\mathbb{I}/2$, which results in a vanishing asymptotic QFI. 
\begin{figure}[htbp]
\centering
\includegraphics[width=1\textwidth]{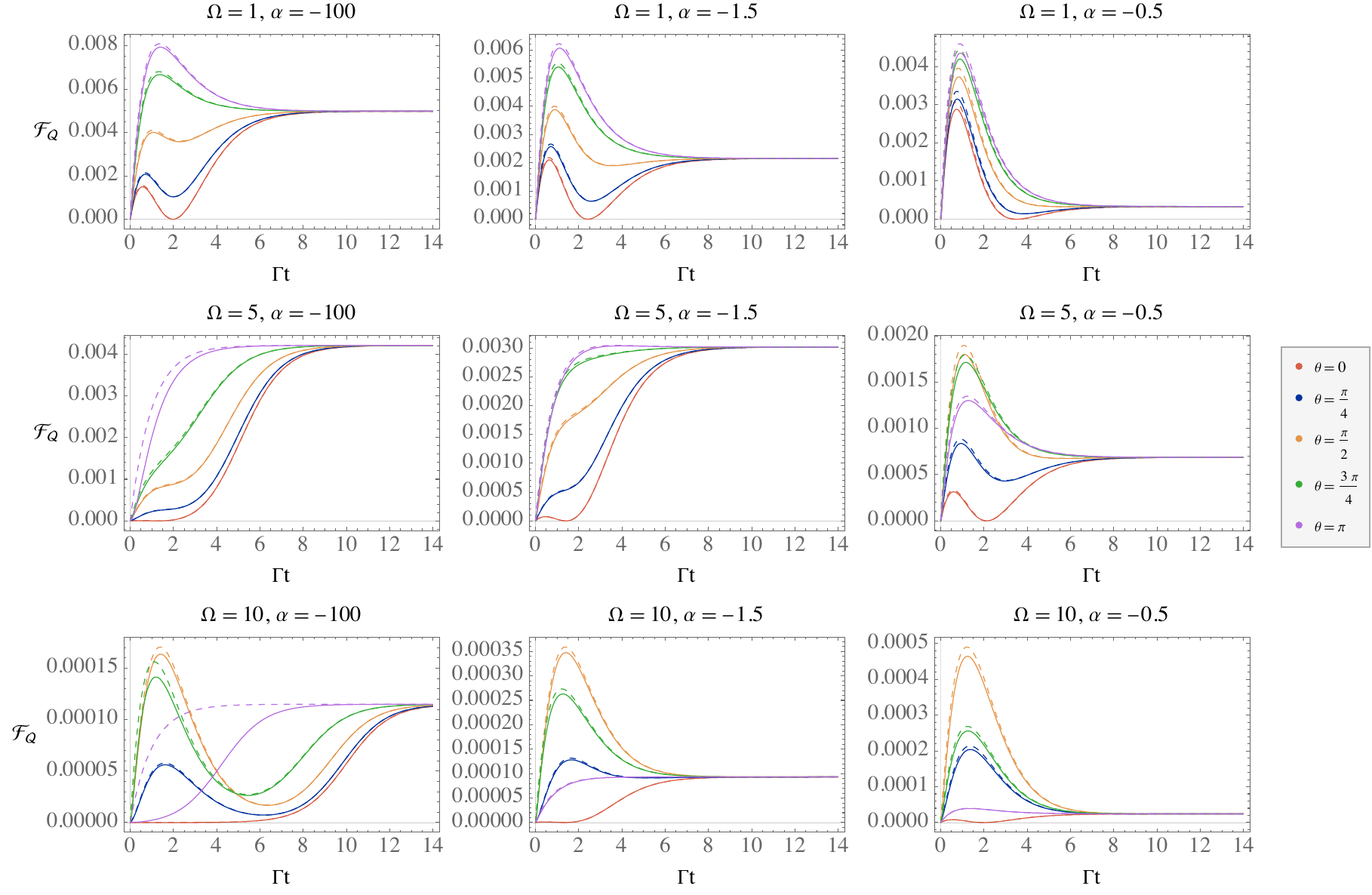}
\caption{Late-time behavior of $\mathcal{F}_{Q}(H;\alpha)$ for Markovian {(dashed lines)} and non-Markovian {(solid lines)} solutions. The estimation has been taken under $g=0.1$, $H=2\pi$ {and $\kappa H=10^{-3}$}, for various de Sitter vacuum-choices ($\alpha=-100,-1.5,-0.5$), detector energy gap (with $\Omega=1,5,10$) and initial states (with $\theta=0,\pi/4,\pi/2,3\pi/4,\pi$). \label{aNMMFL100}}
\end{figure}

Like in the Bunch-Davies case, we can prove the {vanishingness} of asymptotic $\mathcal{F}_{Q}(H;\alpha)$ also at the limits of {extremely} small/large detector energy gap. {For $\Omega\rightarrow0$, we still obtain the $H$-independent equilibrium state $\rho^{non-BD}_{\text{asym}}\approx\mathbb{I}/2$, giving vanishing $\mathcal{F}_{Q}(H;\alpha)$.} For large $\Omega$, we obtain
\begin{equation}
    \lim_{\Omega\rightarrow \infty}\gamma_{(0)}=\lim_{\Omega\rightarrow \infty}-\frac{\sinh{\alpha}\sinh{\left(\frac{\pi\Omega}{H}\right)}}{1+\cosh{\alpha}\cosh{\left(\frac{\pi\Omega}{H}\right)}}=-\tanh{\alpha},
\end{equation}
which turns the detector equilibrium state \eqref{asymptstate-ds} into
\begin{equation}
    \rho^{non-BD}_{\text{asym}}=\frac{1}{2}\begin{bmatrix}
    1+\tanh{\alpha} && 0
    \\
    0 && 1-\tanh{\alpha}
    \end{bmatrix}.
    \label{alphaasym}
\end{equation}
This is an asymptotic state {independent of the Hubble parameter}, leading to vanishing QFI. 

We end this section with an interpretation of \eqref{alphaasym}. For the detector initially in the ground state, it requires an amount of energy $\Delta E=\hbar \Omega$ to be excited to a higher level. For the Bunch-Davies vacuum, we know that the detector eventually arrives the equilibrium state $\rho^{BD}_{\text{asym}}\approx|0\rangle\langle0|/2$ once $\Omega\rightarrow\infty$. This is reasonable since the de Sitter thermal bath at finite-temperature $T_{BD}=H/2\pi$ can not provide enough energy to excite the detector. However, for general $\alpha$-vacua, \eqref{alphaasym} indicates that the detector can be excited from the ground state $|0\rangle$ even when an infinite excitation energy $\Delta E\gg 0$ is required. The only possibility for this can happen is that the background field states $|\alpha\rangle$ should have infinite energy \cite{dS4,alpha-1}. Historically, this phenomenon has sparked years of debate regarding the pathological nature of $\alpha-$states in a realistic cosmology \cite{alpha-2}. Nevertheless, it is interesting to note that such debated pathology has never been diagnosed by QFI as we have done.

\section{Conclusions}
In this work, we have resolved the non-Markovian dynamics of a comoving UDW detector in de Sitter space. In terms of the quantum Fisher information of Hubble parameter estimation, we found that the non-Markovian contribution is very robust and stable at early times. At late times, we showed that the non-Markovian effect generally reduces the QFI, which converges to an asymptotic value independent of the detector's initial state preparation. Moreover, we are particularly interested in the influence of de Sitter $\alpha$-vacuum choices on detector dynamics, as manifested by the related QFI. For general $\alpha$-vacua, we found that the asymptotic values of QFI can be significantly suppressed compared to the case of the Bunch-Davies vacuum. Due to recent evidence of non-Markovianity in inflationary cosmology \cite{Add1}, we hope our result can provide insight into real observations. On the other hand, quantum simulation of de Sitter space quantum fields in the lab \cite{Add2} may also be an interesting test platform of our theory.

The method adopted in the paper can be generalized to other important spacetime geometries, such as a black hole \cite{Open1-5} or an FRW universe. The related QFI is then anticipated to play a similar role in characterizing detector dynamics, as well as specific features of quantum fields in a curved background. On the other hand, once generalized to multi-parameter $X_i$ estimation, the related QFI matrix $\mathcal{F}_{ij}$ can be identified as a metric in state space from an information geometry perspective. It is related to many important concepts in quantum thermodynamics \cite{QFI1-4,QFI1-5}, e.g., relative entropy as $\mathcal{F}_{ij}=\partial^2 D(\mathbf{X}|\mathbf{X}')/\partial X_i\partial X_j$. We expect that such a link can be manifested in the future through a discussion on the quantum thermodynamics of the UDW detector \cite{Open7} in de Sitter space.

\section*{Acknowledgments}
This work is supported by the National Natural Science Foundation of China (No.12475061, 12075178) and Shaanxi Fundamental Science Research Project for Mathematics and Physics (No. 23JSY006).

\appendix
\section{Solving the Laplace-transformed master equation}
\label{appendixA}
In Section \ref{sec:Master Equation}, we obtained the Laplace-transformed non-Markovian master equation \eqref{REDFIELD3}
\begin{equation}
    z\tilde{\rho}(z)-\rho(0)=g^{2}\sum_{\mu=0,\pm}\int_{0}^{\tau}ds\,[\sigma^{(\mu)}\tilde{\rho}(z),\sigma^{\dag}_{(\mu)}]\tilde{\Delta}^{+}(z+i\mu\omega)+h.c.
\end{equation}
From the decomposition $\rho=\frac{1}{2}(1+\sum_{\mu}v_{(\mu)} \sigma^{(\mu)})$, we have
\begin{equation}
    \tilde{\rho}=\frac{1}{2}\left(\frac{1}{z}+\sum_{\mu}\tilde{v}_{(\mu)}\sigma^{(\mu)}\right).
\end{equation}
which further gives
\begin{equation}
    \left[\sigma^{(\mu)}\tilde{\rho}(z),\sigma^{\dag}_{(\mu)}\right]=\frac{1}{2}\left(\frac{1}{z}[\sigma^{(\mu)},\sigma^{\dag}_{(\mu)}]+\sum_{\nu}\tilde{v}_{(\nu)}[\sigma^{(\mu)}\sigma^{(\nu)},\sigma^{\dag}_{(\mu)}]\right).
    \label{rhocommute}
\end{equation}

Using \eqref{n}, we know that 
\begin{equation}
\left[\sigma^{(\mu)},\sigma^{(\nu)}\right]=-2\sum_{\gamma}\epsilon^{\mu\nu\gamma}\sigma^{\dag}_{\gamma},
    ~~~~~~
    \left\{\sigma^{(\mu)},\sigma^{\dag}_{(\nu)}\right\}=2\delta^{\mu}_{\,\nu}
\end{equation}
which leads
\begin{equation}
   \left[\sigma^{(\mu)}\tilde{\rho}(z),\sigma^{\dag}_{(\mu)}\right]=\frac{\mu\sigma^{(0)}}{z}-\sum_{\nu\neq \mu}\tilde{v}_{(\nu)}\sigma^{(\nu)}.
\end{equation}
Substituting it back in the Laplace-transformed Master equation \eqref{rhocommute}, one obtains
\begin{equation}
    \begin{aligned}
        z\tilde{\rho}(z)-\rho(0)=
        g^{2}\left(M(z)\sigma^{(0)}/z-\sum_{\nu=0,\pm}N_{(\nu)}(z)\tilde{v}_{(\nu)}\sigma^{(\nu)}\right),
            \label{A-1}
    \end{aligned}
\end{equation}
where
\begin{equation}
    \begin{aligned}
        &N_{(\nu)}(z)\equiv\sum_{\mu\neq\nu}[\tilde{\Delta}^{+}(z+i\mu\omega)+\tilde{\Delta}^{-}(z+i\mu\omega)],
        \\
        &M(z)\equiv\tilde{\Delta}^{+}(z+i\omega)+\tilde{\Delta}^{-}(z-i\omega)-\tilde{\Delta}^{+}(z-i\omega)-\tilde{\Delta}^{-}(z+i\omega).
    \end{aligned}
\end{equation}
Noticing that
\begin{equation}
    z\tilde{\rho}-\rho(0)=\sum_{\mu=0,\pm}\frac{1}{2}\left[z\tilde{v}_{(\mu)}-v_{(\mu)}(0)\right]\sigma^{(\mu)},
\end{equation}
and comparing its coefficients before $\sigma^{(\mu)}$ with \eqref{A-1}, we obtain the solutions \eqref{v}. 
%\begin{equation}
%    \begin{aligned}
%      \tilde{v}_{(0)}(z)&=\frac{v_{(0)}|_{\tau=0}+2g^{2}M(z)/z}{z+2g^{2}N_{(0)}(z)},\\
%        \tilde{v}_{(\pm)}(z)&=\frac{v_{(\pm)}|_{\tau=0}}{z+2g^{2}N_{(\pm)}(z)},\\
%        \tilde{v}_{0}(p)&=\frac{v_{0}|_{\tau=0}+2g^{2}M(p)/p}{p+2g^{2}N_{0}(p)},\\
%        \tilde{v}_{\pm}(p)&=\frac{v_{\pm}|_{\tau=0}}{p+2g^{2}N_{\pm}(p)}.
%    \end{aligned}
%\end{equation}

\section{Laplace transform of the correlation functions}
\label{appendixB}
In Section \ref{sec:corrlation} we defined the functions $E_{(\pm)}(z)$ and $F(z)$. We start from calculation of $E_{(+)}(z)$:\begin{equation}
    E_{(+)}(z)=-\int_{0}^{\infty}\mathrm{d}\tau\,\frac{e^{-z\tau}}{16\pi^{2}\sinh^{2}((H\tau-i\epsilon)/2)}.
\end{equation}
Applying integration by parts to the denominator, we obtain
\begin{equation}
    E_{(+)}(z)=\frac{H}{8\pi^2}\left(\coth{\frac{i\epsilon}{2}}+\int_{0}^{\infty}\,\mathrm{d}\tau\,ze^{-z\tau}\coth{\frac{H\tau- i\epsilon}{2}}\right).
\end{equation}
Changing the integration variable to $u\equiv e^{-H\tau}$, and expanding $\coth{\frac{H\tau- i\epsilon}{2}}$ as a power series in $u$, we get
\begin{equation}
E_{(+)}(z)=\frac{H}{8\pi^2}\left(\pm\coth{\frac{i\epsilon}{2}}+2z\sum_{n=0}^{\infty}\frac{e^{\pm in\epsilon}}{z+nH}-1\right),
\label{Epm}
\end{equation}
from which $E^{-}(p)$ can also be determined by the substitution $\epsilon\rightarrow-\epsilon$.

In a similar way, we evaluate $F(z)$ as 
\begin{equation}
F(z)=\frac{H}{8\pi^2}\left(2z\sum_{n=0}^{\infty}\frac{(-1)^{n}\cos{n\epsilon}}{z+nH}-1\right).
\label{F}
\end{equation}

%Collecting all above results, we arrive at \eqref{EF}.

It is worth mentioning that although \eqref{Epm} contains linearly divergent terms $\pm\coth{\frac{i\epsilon}{2}}$, they are always canceled when we evaluate $M(z)$ and $N_{(\mu)}(z)$.  On the other hand, the $e^{\pm in\epsilon}$ factors protect the convergence of \eqref{InverseLaplaceTransformation} in an intricate way, and are therefore more important. To make sure that we do everything correctly, we keep $\epsilon$ to be small but finite throughout all calculation, and only neglect contributions that are guaranteed to be small in the {final results} $\rho(t)$. In fact, $\epsilon$ is naturally related to the cutoff of the background fields through $\Lambda/H \sim 1/\epsilon$, which is expected to be a small and positive value.

\section{Calculation of the constants $\gamma_{(0)}$ and $\Gamma_{(\mu)}$}
\label{appdixD}
In this Appendix we aim to calculate the coefficient functions \eqref{coefficient1} and \eqref{coefficient2}, defined by
\begin{equation}
    \begin{aligned}
        \gamma_{(0)}=-\frac{M(0)}{N_{0}(0)},~~~~~~
        \Gamma_{(\mu)}=2g^{2}N_{(\mu)}(0),~~~~~\mu=0,\pm
        \label{eqap1}
    \end{aligned}
\end{equation}
From \eqref{MN}, we know that
\begin{equation}
\begin{aligned}
    M(0;\alpha)&=\tilde{\Delta}^{+}(i\omega;\alpha)+\tilde{\Delta}^{-}(-i\omega;\alpha)-\tilde{\Delta}^{+}(-i\omega;\alpha)-\tilde{\Delta}^{-}(i\omega;\alpha),
    \\
    N_{(0)}(0;\alpha)&=\tilde{\Delta}^{+}(i\omega;\alpha)+\tilde{\Delta}^{-}(-i\omega;\alpha)+\tilde{\Delta}^{+}(-i\omega;\alpha)+\tilde{\Delta}^{-}(i\omega;\alpha),
    \\
    N_{(\pm)}(0;\alpha)&=\tilde{\Delta}^{+}(\mp i\omega;\alpha)+\tilde{\Delta}^{-}(\mp i\omega;\alpha)+\tilde{\Delta}^{+}(0;\alpha)+\tilde{\Delta}^{-}(0;\alpha)
\label{MNN}
\end{aligned}
\end{equation}
Introducing the Fourier-transformed Wightman function $\mathcal{G}(\omega)\equiv\int \mathrm{d}\tau\,e^{i\omega \tau}\Delta^{+}(\tau)$, above coefficient functions can be recasted into
\begin{equation}
\begin{aligned}
    M(0;\alpha)&=\mathcal{G}(-\omega)-\mathcal{G}(\omega),\\
    N_{(0)}(0;\alpha)&=\mathcal{G}(-\omega)+\mathcal{G}(\omega),\\
    \mathfrak{Re}\left[N_{(\pm)}(0;\alpha)\right]&=\mathcal{G}(0)+\frac{\mathcal{G}(-\omega)+\mathcal{G}(\omega)}{2}.
\end{aligned}
\end{equation}

The derivation of $\mathcal{G}(\omega)$ in de Sitter space is standard and straightforwardly \cite{dS12-2,dS10}, by which the coefficient functions become
\begin{equation}
\begin{aligned}
    M(0;\alpha)&=-\frac{\omega}{2\pi}+\mathcal{O}(\epsilon\log{\epsilon}),\\
    N_{(0)}(0;\alpha)&=-\left(\frac{\omega}{2\pi}\right)\frac{1+\cosh{\alpha}\cosh{\left(\frac{\pi\omega}{H}\right)}}{\sinh{\alpha}\sinh{\left(\frac{\pi\omega}{H}\right)}}+\mathcal{O}(\epsilon\log{\epsilon}),
    \label{eqap2}
\end{aligned}
\end{equation}
where the $\mathcal{O}(\epsilon\log{\epsilon})$ term can be estimated through the condition
\begin{equation}
    \sum_{n>0}\frac{|\sin{n\epsilon}|}{n^2}\sim \mathcal{O}(\epsilon\log{\epsilon}),
\end{equation}
which would be proved later in Appendix \ref{sec:estimation}. Substituting \eqref{eqap2} into \eqref{eqap1}, we obtain
\begin{equation}
    \begin{aligned}
        \gamma_{(0)}&=-\frac{\sinh{\alpha}\sinh{\left(\frac{\pi\omega}{H}\right)}}{1+\cosh{\alpha}\cosh{\left(\frac{\pi\omega}{H}\right)}}+\mathcal{O}(\epsilon\log{\epsilon}),\\
        \Gamma_{(0)}&=-\left(\frac{g^{2}\omega}{\pi}\right)\frac{1+\cosh{\alpha}\cosh{\left(\frac{\pi\omega}{H}\right)}}{\sinh{\alpha}\sinh{\left(\frac{\pi\omega}{H}\right)}}+\mathcal{O}(g^{2}\epsilon\log{\epsilon}).
    \end{aligned}
\end{equation}
which is \eqref{coefficient1} as we expected.

To calculate $\Gamma_{(\pm)}$, while its real part is easy to get as $\mathfrak{Re}\left[\Gamma_{(\pm)}\right]=2g^2 \mathfrak{Re}\left[N_{(\pm)}(0;\alpha)\right]$, we need a bit more effort to derive its imaginary part. Based on \eqref{MNN}, \eqref{core} and \eqref{EF}, we have
\begin{equation}
\begin{aligned}
    N_{(\pm)}(0;\alpha)=\pm\frac{i\omega}{2\pi^2}\left(\coth{\alpha}\sum_{n=0}^{\infty}\frac{\cos{n\epsilon}}{nH \mp i\omega}+\csch{\alpha}\sum_{n=0}^{\infty}\frac{(-1)^n \cos{n\epsilon}}{nH \mp i\omega}\right).
\end{aligned}
\end{equation}
Noting that
\begin{equation}
    \sum_{n=0}^{\infty}\frac{\cos{n\epsilon}}{nH \mp i\omega}=-\frac{1}{H}\left(\log{\epsilon}+\psi\left(\mp \frac{i\omega}{H}\right)+\gamma\right)+\mathcal{O}(\epsilon\log{\epsilon}),
\end{equation}
and
\begin{equation}
    \sum_{n=0}^{\infty}\frac{(-1)^{n}\cos{n\epsilon}}{nH-i \omega}=-\frac{1}{2H}\left(\psi\left(\frac{i\omega}{2H}\right)-\psi\left(1+\frac{i\omega}{2H}\right)\right)+\mathcal{O}(\epsilon\log{\epsilon}),
\end{equation}
we eventually obtain \eqref{coefficient2}.
%\begin{equation}
%    \begin{aligned}
%        \Gamma_{(\pm)}=\mp\frac{i\omega g^{2}\left(2\cosh{\alpha}\left[\log{(e^{\gamma}\epsilon)}+\psi(\mp i\omega/H)\right]+\psi\left(\frac{i\omega}{2H}\right)-\psi\left(1+\frac{i\omega}{2H}\right)+\mathcal{O}( \epsilon\log{\epsilon})\right)}{2\pi^{2}\sinh{\alpha}}.
%    \end{aligned}
%\end{equation}

\section{Calculation of the inhomogeneous non-Markovian contributions }
\label{sec:estimation}
In section \ref{sec:Diagonal}, we considered the non-Markovian inhomogeneous contribution controlled by the estimator \eqref{f}, which is
\begin{equation}
\mathfrak{E}(\tau)=\frac{i}{\pi^2}\sum_{n>0}n\sin{n\epsilon}\left(\frac{e^{i\omega\tau-nH\tau}}{(n-i\omega/H)^2}-\frac{e^{-i\omega\tau-nH\tau}}{(n+i\omega/H)^2}\right),
\label{f2}
\end{equation}
Setting $H=1$ for simplicity and expanding $e^{i\omega\tau}$, we get:
\begin{equation}
    \mathfrak{E}(\tau)=-\frac{2}{\pi^2}\sum_{n>0}\sin{n\epsilon}\left(\frac{2n^{2}\omega e^{-n\tau}\cos{\omega\tau}+(n^2-\omega^2)n e^{-n\tau}\sin{\omega\tau}}{(n^2+\omega^2)^2}\right).
\end{equation}
For $\tau\in[0,\infty)$ and $|\sin{\omega\tau}|\leq \omega\tau$, the following inequality holds
\begin{equation}
    |ne^{-n\tau}\sin{\omega\tau}|\leq n\omega\tau e^{-n\tau}\leq \omega.
\end{equation}
which with a little bit of algebra, gives
\begin{equation}
    |\mathfrak{E}(\tau)|\leq \frac{8\omega}{\pi^2}\sum_{n>0}\frac{|\sin{n\epsilon}|}{n^2}
\end{equation}
For given positive number $N>1$, we have
\begin{equation}
    \left|\mathfrak{E}(\tau)\right|\leq \frac{8\omega}{\pi^2}\left(\sum_{n=1}^{N}\frac{|\sin{n\epsilon}|}{n^2}+\sum_{n=N+1}^{\infty}\frac{1}{n^2}\right).
\end{equation}
Using the fact that $|\sin{n\epsilon}|<n\epsilon$ for $n>0$, we must have
\begin{equation}
    |\mathfrak{E}(\tau)|\leq \frac{8\omega}{\pi^2}\left(\epsilon(1+\log{N})+\frac{1}{N}\right).
\end{equation}
Choosing $N=\lceil 1/\epsilon\rceil$, we eventually have
\begin{equation}
    |\mathfrak{E}(\tau)|\leq \frac{8\omega\epsilon\left(3-\log{\epsilon}\right)}{\pi^2},
\end{equation}
which justifies our assertion that $f(\tau)\sim O(\epsilon\log{\epsilon})$ used in Section \ref{sec:Diagonal}.

\section{Details about the renormalization of non-Markovian evolution}
\label{appendixE}

Here we provide the detailed derivation of the renormalization of non-Markovian evolutions of the qubit \eqref{Re_non_Markov}. Similar to the case of a Markovian evolution, we may first renormalize the frequency $\omega$ and express the result in terms of  $\Omega$ only. In Schr\"{o}dinger picture we have (up to $\mathcal{O}(g^4)$ error)

\be
\left\{\begin{aligned}
v_{(0)}(\tau)   &=\gamma_{(0)}^P\left(e^{-\Gamma_{(0)}^P\tau}-1\right)+v_{(0)}(0)\Big[1+g^2\left(S_{(0)}(\tau,\Omega)-S_{(0)}(0,\Omega)\right)\Big]e^{-\Gamma_{(0)}^P\tau}, \\
v_{(\pm)}(\tau)&=v_{(\pm)}(0)\Big[1+g^2\left(S_{(\pm)}(\tau,\Omega)-S_{(\pm)}(0,\Omega)\right)\Big]e^{-\Gamma_{(\pm)}^P \tau}.
\end{aligned}
\right.
\ee
For simplicity, let's focus on the diagonal part. Setting $\tau=\kappa$ we obtain 
\begin{equation}
    v_{(0)}(\kappa)   =\gamma_{(0)}^P\left(e^{-\Gamma_{(0)}^P\kappa}-1\right)+v_{(0)}(0)\Big[1+g^2\left(S_{(0)}(\kappa,\Omega)-S_{(0)}(0,\Omega)\right)\Big]e^{-\Gamma_{(0)}^P\kappa},
\end{equation}
Solving $v_{(0)}(0)$ in terms of  $v_{(0)}(\kappa)$ we have
\begin{equation}
    v_{(0)}(0)=\frac{e^{\Gamma_{(0)}^P\kappa}\left(v_{(0)}(\kappa)-\gamma_{(0)}^P\left(e^{-\Gamma_{(0)}^P\kappa}-1\right)\right)}{1+g^2\left(S_{(0)}(\kappa,\Omega)-S_{(0)}(0,\Omega)\right)}.
\end{equation}
Choosing the new time variable $t=\tau-\kappa$, we obtain 
\begin{equation}
    v_{(0)}(t)   =\gamma_{(0)}^P\left(e^{-\Gamma_{(0)}^Pt}-1\right)+v_{(0)}(\kappa)\Big[1+g^2\left(S_{(0)}^P(t,\Omega)-S_{(0)}^P(0,\Omega)\right)\Big]e^{-\Gamma_{(0)}^Pt}+\mathcal{O}(g^4),
\end{equation}
where we have also defined $S^P(t)\equiv S(t+\kappa)$ for simplicity. 

The treatment of off-diagonal elements is similar. Combining the above results together we have:
\be
\left\{\begin{aligned}
v_{(0)}(t)   &=\gamma_{(0)}^P\left(e^{-\Gamma_{(0)}^Pt}-1\right)+v_{(0)}(0)\Big[1+g^2\left(S_{(0)}^P(t,\Omega)-S_{(0)}^P(0,\Omega)\right)\Big]e^{-\Gamma_{(0)}^Pt}, \\
v_{(\pm)}(t)&=v_{(\pm)}(0)\Big[1+g^2\left(S_{(\pm)}^P(t,\Omega)-S_{(\pm)}^P(0,\Omega)\right)\Big]e^{-\Gamma_{(\pm)}^P t}.
\label{Re_non_Markov_2}
\end{aligned}
\right.
\ee
Since $\kappa$ is a finite physical scale that is independent of the cut-off $\epsilon$, the expression \eqref{Re_non_Markov_2} is free of divergences and we may take $\epsilon=0$ in $S^P_{(\mu)}(t)$.


\begin{thebibliography}{999}

\bibitem{DS1}
D. Anninos, \textit{De Sitter musings}, \href{https://www.worldscientific.com/doi/abs/10.1142/S0217751X1230013X}{Int. J. Mod. Phys. A \textbf{27}, 1230013 (2012)}.


\bibitem{dS1}
E. Witten, \textit{Quantum Gravity In De Sitter Space},  \href{https://doi.org/10.48550/arXiv.hep-th/0106109}{arXiv:hep-th/0106109}.


\bibitem{UDW1}
W. G. Unruh, \textit{Notes on black-hole evaporation}, \href{https://journals.aps.org/prd/abstract/10.1103/PhysRevD.14.870}{Phys. Rev. D \textbf{14}, 870 (1976)}. 

\bibitem{UDW2}
B. S. DeWitt, Quantum gravity: the new synthesis, in \textit{General Relativity: An Einstein Centenary Survey}, edited by S.W. Hawking and W. Israel (Cambridge University Press, Cambridge, 1979), pp. 680.

\bibitem{UDW3}
N. D. Birrell and P. C. W. Davies, \textit{Quantum fields in curved space}, Cambridge University Press (1982).

\bibitem{UDW4}
L. C. B. Crispino, A. Higuchi and G. E. A. Matsas, \textit{The Unruh effect and its applications}, \href{https://journals.aps.org/rmp/abstract/10.1103/RevModPhys.80.787}{Rev. Mod. Phys. \textbf{80}, 787 (2008)}.

\bibitem{UDW5}
P. Candelas, \textit{Vacuum polarization in Schwarzschild spacetime}, \href{https://journals.aps.org/prd/abstract/10.1103/PhysRevD.21.2185}{Phys. Rev. D \textbf{21}, 2185 (1980)}.

\bibitem{UDW6}
S. W. Hawking, \textit{Particle creation by black holes}, \href{https://link.springer.com/article/10.1007/BF02345020}{Commun. Math. Phys. \textbf{43}, 199 (1975)}.

\bibitem{dS12-1}
G. W. Gibbons and S. W. Hawking, \textit{Cosmological event horizons, thermodynamics, and particle creation}, \href{https://journals.aps.org/prd/abstract/10.1103/PhysRevD.15.2738}{Phys. Rev. D \textbf{15}, 2738 (1977)}.

\bibitem{dS12-2}
B. Garbrecht and T. Prokopec, \textit{Unruh response functions for scalar fields in de Sitter space}, \href{https://iopscience.iop.org/article/10.1088/0264-9381/21/21/016/pdf}{Class. Quant. Grav. \textbf{21}, 4993 (2004)}.

\bibitem{UDW7}
G. L. Sewell, \textit{Quantum fields on manifolds: PCT and gravitationally induced thermal states}, \href{https://www.sciencedirect.com/science/article/abs/pii/0003491682902858}{Ann. Phys.
\textbf{141}, 201 (1982)}.

\bibitem{UDW8}
S. Takagi, \textit{Vacuum Noise and Stress Induced by Uniform Acceleration}, \href{https://doi.org/10.1143/PTP.88.1}{Prog. Theor. Phys. Suppl. \textbf{88}, 1 (1986)}.

\bibitem{UDW9}
J. Arrechea, C. Barcel\'o, L. J. Garay, and G. Garc\'ia-Moreno, \textit{Inversion of statistics and thermalization in the Unruh effect}, \href{https://journals.aps.org/prd/abstract/10.1103/PhysRevD.104.065004}{Phys. Rev. D \textbf{104}, 065004 (2021)}.

\bibitem{Open1-1}
F. Benatti and R. Floreanini, \textit{Entanglement generation in uniformly accelerating atoms: reexamination of the Unruh effect}, \href{https://journals.aps.org/pra/abstract/10.1103/PhysRevA.70.012112}{Phys. Rev. A \textbf{70}, 012112 (2004)}.

\bibitem{Open1-2}
H. Yu and J. Zhang, \textit{Understanding Hawking radiation in the framework of open quantum systems}, \href{https://journals.aps.org/prd/abstract/10.1103/PhysRevD.77.024031}{Phys. Rev. D \textbf{77}, 024031 (2008)}.

\bibitem{Open1-3}
H. Yu, \textit{Open quantum system approach to Gibbons-Hawking effect of de Sitter space-time}, \href{https://journals.aps.org/prl/abstract/10.1103/PhysRevLett.106.061101}{Phys. Rev. Lett. \textbf{106}, 061101 (2011)}.

\bibitem{Open1-4}
G. Kaplanek and C.P. Burgess, \textit{Hot accelerated qubits: decoherence, thermalization, secular growth and reliable late-time predictions}, \href{https://link.springer.com/article/10.1007/JHEP03(2020)008}{J. High Energy Phys. \textbf{03} (2020) 008}.

\bibitem{Open1-5}
G. Kaplanek and C.P. Burgess, \textit{Qubits on the horizon: decoherence and thermalization near black holes}, \href{https://link.springer.com/article/10.1007/JHEP01(2021)098}{J. High Energy Phys. \textbf{01} (2021) 098}.



\bibitem{Open2}
J. Feng, Y. -Z. Zhang, M. D. Gould, and H. Fan, \textit{Uncertainty relation in Schwarzschild spacetime}, \href{https://www.sciencedirect.com/science/article/pii/S0370269324004349}{Phys. Lett. B \textbf{743}, 198 (2015)}.

\bibitem{Open3}
L. Jia, Z. Tian, and J. Jing, \textit{Entropic uncertainty relation in de Sitter space}, \href{https://www.sciencedirect.com/science/article/abs/pii/S0003491614003133}{Ann. Phys. \textbf{353}, 37 (2015)}.

\bibitem{Open4}
J. Hu and H. Yu, \textit{Geometric phase outside a Schwarzschild black hole and the Hawking effect}, \href{https://link.springer.com/article/10.1007/JHEP09(2012)062}{J. High Energy Phys. \textbf{09} (2012) 062}.

\bibitem{Open5}
Z. Tian and J. Jing, \textit{Geometric phase of two-level atoms and thermal nature of de Sitter spacetime}, \href{https://link.springer.com/article/10.1007/JHEP04(2013)109}{J. High Energy Phys. \textbf{04} (2013) 109}.

\bibitem{Open6}
J. Feng, J.-J. Zhang, and Y. Zhou, \textit{Thermality of the Unruh effect with intermediate statistics}, \href{https://iopscience.iop.org/article/10.1209/0295-5075/ac5ddb/pdf}{Europhys. Lett. \textbf{137}, 60001 (2022)}.

\bibitem{Open7}
S.-W. Han, Z. Ouyang, Z. Hu, and J. Feng, \textit{Relative entropy formulation of thermalization process in a Schwarzschild spacetime}, \href{https://www.sciencedirect.com/science/article/pii/S0370269324007937}{Phys. Lett. B \textbf{861} (2025) 139235}.


\bibitem{dS12}
M. Fukuma, S. Sugishita, and Y. Sakatani, \textit{Master equation for the Unruh-DeWitt detector and the universal relaxation time in de Sitter space}, \href{https://journals.aps.org/prd/abstract/10.1103/PhysRevD.89.064024}{Phys. Rev. D \textbf{89}, 064024 (2014)}.

\bibitem{dS13}
G. Kaplanek and C.P. Burgess, \textit{Hot cosmic qubits: late-time de Sitter evolution and critical slowing down}, \href{https://link.springer.com/article/10.1007/JHEP02(2020)053}{J. High Energ. Phys \textbf{02} (2020) 053}.

\bibitem{dS14}
C.P. Burgess, R. Holman, G. Kaplanek, J. Marting and V. Vennin, \textit{Minimal decoherence from inflation}, \href{https://iopscience.iop.org/article/10.1088/1475-7516/2023/07/022}{J. Cosmol. Astropart. \textbf{07} (2023) 022}.




\bibitem{Open1}
H.-P. Breuer and F. Petruccione, \textit{The Theory of Open Quantum Systems} (Oxford University Press 2002).


\bibitem{dS15-1}
G. Kaplanek and E. Tjoa, \textit{Effective master equations for two accelerated qubits}, \href{https://journals.aps.org/pra/abstract/10.1103/PhysRevA.107.012208}{Phys. Rev. A \textbf{107}, 012208 (2023)}.

\bibitem{dS15-2}
D. Moustos and C. Anastopoulos, \textit{Non-Markovian time evolution of an accelerated qubit}, \href{https://journals.aps.org/prd/abstract/10.1103/PhysRevD.95.025020}{Phys. Rev. D \textbf{95}, 025020 (2017)}.

\bibitem{dS15-3}
D. Moustos, \textit{Asymptotic states of accelerated detectors and universality of the Unruh effect}, \href{https://journals.aps.org/prd/abstract/10.1103/PhysRevD.98.065006}{Phys. Rev. D \textbf{98}, 065006 (2018)}.

\bibitem{dS2}
E. Mottola, \textit{Particle creation in de Sitter space}, \href{https://journals.aps.org/prd/abstract/10.1103/PhysRevD.31.754}{Phys. Rev. D \textbf{31}, 754 (1985)}.

\bibitem{dS3}
B. Allen, \textit{Vacuum states in de Sitter space}, \href{https://journals.aps.org/prd/abstract/10.1103/PhysRevD.32.3136}{Phys. Rev. D \textbf{32}, 3136 (1985)}.

\bibitem{dS4}
M. B. Einhorn and F. Larsen, \textit{Interacting quantum field theory in de Sitter vacua}, \href{https://journals.aps.org/prd/abstract/10.1103/PhysRevD.67.024001}{Phys. Rev. D \textbf{67}, 024001 (2003)}.

\bibitem{dS5}
K. Goldstein and D. A. Lowe, \textit{A note on $\alpha$-vacua and interacting field theory in de Sitter space}, \href{
https://doi.org/10.1016/j.nuclphysb.2003.07.014}{Nucl. Phys. B \textbf{669}, 325 (2003)}.

\bibitem{dS6}
H. Collins, R. Holman, and M. R. Martin, \textit{The fate of the alpha-vacuum}, \href{https://journals.aps.org/prd/abstract/10.1103/PhysRevD.68.124012}{Phys. Rev. D \textbf{68}, 124012 (2003)}.

\bibitem{dS7}
J. de Boer, V. Jejjala, and D. Minic, \textit{$\alpha$-states in de Sitter space}, \href{https://journals.aps.org/prd/abstract/10.1103/PhysRevD.71.044013}{Phys. Rev. D \textbf{71}, 044013 (2005)}.

\bibitem{dS8}
U. H. Danielsson, \textit{Inflation, holography, and the choice of vacuum in de Sitter space}, \href{https://doi.org/10.1088/1126-6708/2002/07/040}{J. High Energy Phys. \textbf{07} (2002) 040}.

\bibitem{dS8+}
U. H. Danielsson, \textit{Note on inflation and trans-Planckian physics}, \href{https://journals.aps.org/prd/abstract/10.1103/PhysRevD.66.023511}{Phys. Rev. D \textbf{66}, 023511 (2002)}.

\bibitem{dS9}
K. Goldstein and D. A. Lowe, \textit{Quantum initial conditions for inflation and canonical invariance}, \href{https://journals.aps.org/prd/abstract/10.1103/PhysRevD.102.023507}{Phys. Rev. D \textbf{69}, 023507 (2004)}.

\bibitem{dS10}
R. Bousso, A. Maloney and A. Strominger, \textit{Conformal vacua and entropy in de Sitter space}, \href{https://journals.aps.org/prd/abstract/10.1103/PhysRevD.65.104039}{Phys. Rev. D \textbf{65}, 104039 (2002)}.

\bibitem{dS11}
U. Danielsson, \textit{The quantum swampland}, \href{https://link.springer.com/article/10.1007/JHEP04(2019)095}{J. High Energy Phys. \textbf{04}, 095 (2019)}.

\bibitem{Open8-1}
 M. Aspachs, G. Adesso, and I. Fuentes, \textit{Optimal Quantum Estimation of the Unruh-Hawking Effect}, \href{https://journals.aps.org/prl/abstract/10.1103/PhysRevLett.105.151301}{Phys. Rev. Lett. \textbf{105}, 151301 (2010)}.

\bibitem{Open8-2}
Z. Tian, J. Wang, H. Fan, and J. Jing, \textit{Relativistic Quantum Metrology in Open System Dynamics}, \href{https://www.nature.com/articles/srep07946}{Sci. Rep. \textbf{5}, 7946 (2015)}.

\bibitem{Open8-3}
J. Feng and J.-J. Zhang, \textit{Quantum Fisher information as a probe for Unruh thermality}, \href{https://doi.org/10.1016/j.physletb.2022.136992}{Phys. Lett. B \textbf{827}, 136992 (2022)}.

\bibitem{Open8-4}
J. Wang, Z. Tian, J. Jing, and H. Fan, \textit{Parameter estimation for an expanding universe}, \href{https://www.sciencedirect.com/science/article/pii/S0550321315000322}{Nucl. Phys. B \textbf{892}, 390 (2015)}.

\bibitem{Open8-6}
H. Du and R. B. Mann, \textit{Fisher information as a probe of spacetime structure: Relativistic quantum metrology in (A)dS}, \href{https://link.springer.com/article/10.1007/JHEP05(2021)112}{J. High Energ. Phys \textbf{05} (2021) 112}.

\bibitem{Open8-7}
E. Pattersona and R. B. Mann, \textit{Fisher information of a black hole spacetime}, \href{https://link.springer.com/article/10.1007/JHEP06(2023)214}{J. High Energ. Phys. \textbf{06} (2023) 214}.

\bibitem{Open8-5}
X. Huang, J. Feng, Y.Z. Zhang, and H. Fan, \textit{Quantum estimation in an expanding spacetime}, \href{https://www.sciencedirect.com/science/article/abs/pii/S0003491618302331}{Ann. Phys. \textbf{397}, 336 (2018)}.

\bibitem{R-1}
H.-P. Breuer, E.-M. Laine, J. Piilo, and B. Vacchini, \textit{Colloquium: Non-Markovian dynamics in open quantum systems}, \href{https://journals.aps.org/rmp/abstract/10.1103/RevModPhys.88.021002}{Rev. Mod. Phys. \textbf{88}, 021002 (2016)}.

\bibitem{Master1}
E. B. Davies, \textit{Markovian Master Equations}, \href{https://link.springer.com/article/10.1007/BF01608389}{Commun. Math. Phys. \textbf{39}, 91 (1974)}.

\bibitem{dS12}
J. Feng, X. Huang, Y.-Z. Zhang, and H. Fan, \textit{Bell inequalities violation within non-Bunch-Davies states}, \href{https://www.sciencedirect.com/science/article/pii/S0370269318307871}{Phys. Lett. B \textbf{786}, 403 (2018)}.

\bibitem{NIST}
F. W. J. Olver, D. W. Lozier, R. F. Boisvert, and C. W. Clark, \textit{NIST Handbook of Mathematical Functions} (Cambridge University Press 2010), pp. 612.

\bibitem{QFI1-1}
C. W. Helstrom, \textit{Quantum detection and estimation theory}, \href{https://link.springer.com/article/10.1007/BF01007479}{J. Stat. Phys. \textbf{1}, 231 (1969)}.

\bibitem{QFI1-2}
S. L. Braunstein and C. M. Caves, \textit{Statistical distance and the geometry of quantum states}, \href{https://journals.aps.org/prl/abstract/10.1103/PhysRevLett.72.3439}{Phys. Rev. Lett. \textbf{72}, 3439 (1994)}.

\bibitem{QFI1-3}
M. G. A. Paris, \textit{Quantum estimation for quantum technology}, \href{https://doi.org/10.1142/S0219749909004839}{Int. J. Quantum Inf. \textbf{7}, 125 (2009)}.

\bibitem{Toolbox}
G. B. Arfken and H. J. Weber, \textit{Mathematical Methods for Physicists}, 6th Edition (Academic Press 2005).

\bibitem{dS15}
M. Spradlin and A. Volovich, \textit{Vacuum states and the S matrix in dS/CFT}, \href{https://journals.aps.org/prd/abstract/10.1103/PhysRevD.65.104037}{Phys. Rev. D \textbf{65}, 104037 (2002)}.

\bibitem{dS16}
S. Kanno, J. Murugan, J. P. Shock and J. Soda, \textit{Entanglement entropy of $\alpha$-vacua in de Sitter space}, \href{https://link.springer.com/article/10.1007/JHEP07(2014)072}{J. High Energ. Phys. \textbf{07} (2014) 072.}

\bibitem{dS17}
S. Ryu and T. Takayanagi, \textit{Holographic derivation of entanglement entropy from AdS/CFT}, \href{https://journals.aps.org/prl/abstract/10.1103/PhysRevLett.96.181602}{Phys. Rev. Lett. 96 (2006) 181602}.

\bibitem{dS18}
R. H. Brandenberger and J. Martin, \textit{Trans-Planckian issues for inflationary cosmology}, \href{https://iopscience.iop.org/article/10.1088/0264-9381/30/11/113001}{Class. Quantum Grav. \textbf{30} (2013) 113001}.

\bibitem{alpha-0}
R. H. Brandenberger, \textit{Initial conditions for inflation — A short review}, \href{https://doi.org/10.1142/S0218271817400028}{Int. J. Mod. Phys. D \textbf{26}, 1740002 (2017)}.

\bibitem{alpha-1}
N. Kaloper, M. Kleban, A. Lawrence, S. Shenker and L. Susskind, \textit{Initial conditions for inflation}, \href{https://doi.org/10.1088/1126-6708/2002/11/037}{J. High Energy Phys. \textbf{11}, 037 (2002)}.

\bibitem{alpha-2}
U. Danielsson, \textit{On the consistency of de Sitter vacua}, \href{https://doi.org/10.1088/1126-6708/2002/12/025}{J. High Energy Phys. \textbf{12}, 025 (2012)}.

\bibitem{alpha-3}
A. Shukla, S. P. Trivedi and V. Vishal, \textit{Symmetry constraints in inflation, $\alpha$-vacua, and the three point function}, \href{https://link.springer.com/article/10.1007/JHEP12(2016)102}{J. High Energy Phys. \textbf{12}, 102 (2016)}.

\bibitem{alpha-4}
P. A. R. Ade, \emph{et al.}, Planck Collaboration, \textit{Planck 2015 results XX. Constraints on inflation}, \href{https://doi.org/10.1051/0004-6361/201525898}{Astron. Astrophys. \text{594} (2016) A20}.


\bibitem{NonM1}
H. P. Breuer, E. Laine, J. Piilo, and B. Vacchini, \textit{Non-Markovian dynamics in open quantum systems}, \href{https://doi.org/10.1103/RevModPhys.88.021002}{Rev. Mod. Phys. \textbf{88}, 021002 (2016)}.

\bibitem{Add1}
S. Brahma, A. Berera and J. Calder\'on-Figueroa, \textit{Quantum corrections to the primordial tensor spectrum: open EFTs \& Markovian decoupling of UV modes}, \href{https://link.springer.com/article/10.1007/JHEP08(2022)225}{J. High Energy Phys. \textbf{08}, 225 (2022).}

\bibitem{Add2}
P. O. Fedichev and U. R. Fischer, \textit{Gibbons-Hawking Effect in the Sonic de Sitter Space-Time of an Expanding Bose-Einstein-Condensed Gas}, \href{https://doi.org/10.1103/PhysRevLett.91.240407}{Phys. Rev. Lett. \textbf{91}, 240407 (2003).}

\bibitem{QFI1-4}
J. S. Sidhu and P. Kok, \textit{Geometric perspective on quantum parameter estimation}, \href{https://doi.org/10.1116/1.5119961}{AVS Quantum Sci. \textbf{2}, 014701 (2020)}.

\bibitem{QFI1-5}
J. Lambert and E. S. S\o rensen, \textit{From classical to quantum information geometry: a guide for physicists}, \href{https://doi.org/10.1088/1367-2630/aceb14}{New J. Phys. \textbf{25}, 081201 (2023)}.



\end{thebibliography}
\end{document}